\documentclass[useAMS,usenatbib,psfig]{mn2e}

\usepackage{graphicx}
\usepackage{lscape}
\usepackage{rotating}
\usepackage{pst-all}
\usepackage{amssymb}

\newcommand{\etal}{{\it et~al.}}

\newcommand{\IRAS}{{\it IRAS}}

\newcommand{\uchii}{{UC H{\scriptsize II}}}

\newcommand{\mstar}{M$_\odot$}
\newcommand{\lum}{L$_\odot$} 
 
\newcommand{\kms}{\mbox{kms$^{-1}$}}
\newcommand{\nh}{\mbox{NH$_3$}}

\newcommand{\nhone}{\mbox{NH$_3$(1,1)}}
\newcommand{\nhtwo}{\mbox{NH$_3$(2,2)}}

\newcommand{\vlsr}{\mbox{V$_{\rm LSR}$}}
\newcommand{\delv}{\mbox{$\Delta$V}}
\newcommand{\trot}{\mbox{T$_{\rm rot}$}}
\newcommand{\tkin}{\mbox{T$_{\rm kin}$}}

\title[]{Physical characterisation of southern massive star-forming regions using Parkes \nh\, observations}
\author[T. Hill \etal]{T. Hill$^{1}$\thanks{E-mail:thill@astro.ex.ac.uk}, S. N. Longmore$^{2}$,  C. Pinte$^1$,  M. R. Cunningham$^3$, M. G. Burton$^3$ \and and V. Minier$^{4,5}$\\
$^{1}$ School of Physics, University of Exeter, Stocker Road, EX4 4QL, Exeter, UK\\
$^{2}$Harvard-Smithsonian Center for Astrophysics, 60 Garden Street, Cambridge, MA 02138, USA\\
$^3$ School of Physics, University of New South Wales, Sydney, 2052, NSW, Australia.\\
$^4$ CEA, DSM, IRFU, Service d'Astrophysique, 91191 Gif-sur-Yvette, France\\
$^5$ Laboratoire AIM, CEA/DSM - CNRS - Universit\'e Paris Diderot, IRFU/Service d'Astrophysique, CEA-Saclay, 91191 \\Gif-sur-Yvette, France 
}

\begin{document}

\date{Accepted/Received}

\pagerange{\pageref{firstpage}--\pageref{lastpage}} \pubyear{2009}

\maketitle

\label{firstpage}

\begin{abstract}

We have undertaken a Parkes ammonia spectral line study, in the lowest two inversion transitions, of southern massive star formation regions, including young massive candidate protostars, with the aim of characterising the earliest stages of massive star formation. 138 sources from the submillimetre continuum emission studies of Hill \etal\, were found to have robust (1,1) detections, including two sources with two velocity components, and 102 in the (2,2) transition.

We determine the ammonia line properties of the sources: linewidth, flux density, kinetic temperature, \nh\, column density and opacity, and revisit our SED modelling procedure to derive the mass for 52 of the sources. By combining the continuum emission information with ammonia observations we substantially constrain the physical properties of the high-mass clumps. There is clear complementarity between ammonia and continuum observations for derivations of physical parameters. 

The MM-only class, identified in the continuum studies of Hill \etal, display smaller sizes, mass and velocity dispersion and/or turbulence than star-forming clumps, suggesting a quiescent prestellar stage and/or the formation of less massive stars.
\end{abstract}


\section{Introduction}\label{sec:intro}

Massive stars are dynamical and enigmatic powerhouses that shape and drive both their local stellar neighbourhood and the ecology and evolution of their host galaxy. Despite this heavy influence, the formation and evolution of a massive star is not well understood. Particularly perplexing are the earliest evolutionary stages of massive star formation (MSF). The difficulty lies in the unambiguous detection, identification and characterisation of these stages, snapshots of which are difficult to obtain as a result of the general rarity of candidates and the rapidity of their evolution \citep{garay99}. 

Whether or not massive stars are scaled-up analogs of low-mass stars is still uncertain. Massive stars exert considerable radiative pressure on the surrounding dust and gas, which in principle could halt and reverse spherical infall of the collapsing protostar. Therefore, a `simple' scaled-up version of low mass star formation is insufficient for massive stars. There are a number of different approaches to address this dilemma as outlined in the reviews by \citet{evans02, menten05, zinnecker07, beuther07}.

As their sophistication increases numerical simulations in two and three dimensions show that the radiation pressure problems associated with spherical symmetry can be overcome \citep[e.g.][]{krumholz09}. It is still not clear however, whether the bulk of the mass that finally ends up on the massive star comes from the monolithic collapse of a single dense core \citep[e.g.][]{mckee03} or is accreted from surrounding lower density gas which is funneled to the centre of a cluster's gravitational potential where the most massive stars are forming \citep[e.g][]{bonnell98}.

The natal molecular cloud, from which high-mass stars are formed, is expected to be dense, massive and cold, detectable only at (sub)millimetre wavelengths. Within the molecular cloud, core collapse will be triggered. As the collapsing protostar gains mass, the gravitational energy will serve to heat the core and ionise the surrounding material, causing an increase in the luminosity of the core. In terms of parameter evolution, the initial protostar will be massive, cool and of low luminosity. As the core evolves, it will accumulate more mass, and the temperature and luminosity of the core will increase. This evolution is seen in low mass protostars \citep{evans02}.

In a search for cold cores that would mark the earliest stages of massive star formation \citet{hill05} undertook a (1.2) millimetre SIMBA\footnote{The SEST IMaging Bolometer Array (SIMBA) was a 37 channel hexagonal array on the Swedish ESO Submillimetre Telescope (SEST).} continuum emission study of sources exhibiting signs of methanol maser and/or radio continuum emission - both of which have previously proven successful tracers of the earliest stages of massive star formation \citep[e.g.][]{minier00, williams04, pestalozzi05}. This SIMBA survey revealed a large number of millimetre continuum sources (255) devoid of the maser and \uchii\, sources targeted. Subsequent submillimetre observations of these sources, which were dubbed `MM-only cores', unveiled submillimetre equivalents for each - often revealing the multiplicity of individual MM-only sources, and confirmed their association with cold, deeply embedded objects \citep{hill06}.

The aforementioned SIMBA survey detected a total of 405 millimetre continuum sources, which (based on their star formation association) could be broken into four classes of source. \citet{hill05} proposed that each of these classes of source could represent a different phase of massive star evolution, with the MM-only class a possible example of the very earliest stages of massive star formation.  A caveat however, is whether the MM-only cores are currently undergoing or will/can support massive star formation.

In order to determine the characteristic properties of the sources in the SIMBA sample, in particular the previously unknown and unstudied MM-only cores, \citet{hill09} performed spectral energy distribution (SED) modelling of a significant fraction of the sample.  SED diagrams are useful tools from which physical quantities such as the luminosity, mass and temperature can be derived \citep[cf.][]{whitney03, robitaille06, hill09}. If the observational data are well constrained \citep[cf.][]{minier05}, SED modelling can provide useful estimates of each of these parameters, for each star-forming core, which allows estimation of  the evolutionary phase of a young massive star.

The luminosity, mass and temperature of an astronomical source are fundamental physical attributes that may be used to clarify and characterise the nature of a source, possibly providing insight into their evolutionary status similarly as they do for low mass stars \citep{andre00}. Indeed, the temperature of a source places important physical constraints on the chemical composition of the core, including which chemical species are present in the core, as well as the size and types of grains present in the central star-forming cores.

The method employed for the SED fitting was that of Bayesian inference, which enabled a statistically probable range of suitable values for the luminosity, mass and temperature, for each source modelled. As SED modelling is heavily reliant upon robust data, it was not possible to usefully constrain each of these parameters for a thorough assessment of massive star formation scenarios. In the absence of reliable far-infrared data, which would serve to constrain the peak of the SED and thus parameters resultant from the fit, additional means of constraining the source parameters, such as temperature, are necessary.

Ammonia is an excellent molecular cloud thermometer \citep{menten05} from which the rotational temperature and ultimately the kinetic temperature of the source can be determined. Ammonia is readily detectable in quiescent dark clouds and regions of low luminosity \citep{ho83} making it perfectly suited to study the birthplace of massive stars. The ratio of the hyperfine components of ammonia's signature five-fingered spectrum provides the optical depth information of the molecular cloud environment. As \nh\, is a particularly good probe of high density gas, and is more resilient to the effects of depletion than other high density tracers (e.g. CS), ammonia observations also provide information pertaining to the density of the core \citep[cf.][and references therein]{longmore07}.

We have undertaken an ammonia spectral line survey of sources from the SIMBA sample of \citet{hill05} in order to obtain an independent determination of their temperature and simultaneously examine the robustness of our SED fitting \citep{hill09}. In addition to facilitating and constraining further SED fitting, these ammonia observations provide information pertaining to the molecular abundance of ammonia in the cores, as well as the physical parameters such as linewidth, rest velocity, column density and virial mass. With this information, we seek to address formation scenarios for massive stars and identify the star formation status/phase of the MM-only core.

%__________________________________________________________________
\section{Observations}\label{sec:obs}

We have undertaken a multi-inversion transition study of ammonia, in the lowest two inversion transitions, toward the SIMBA sample of \citet{hill05}, with the aim of determining characteristic source properties, such as temperature and column density. The sample comprises four classes of millimetre continuum sources as alluded to in section \ref{sec:intro}: class M are sources with a methanol maser association, class R have radio continuum associations, class MR have both a maser and radio continuum source, whilst class MM are detected solely from their millimetre continuum emission.

The ammonia observations were undertaken on the Australia Telescope National Facility (ATNF) operated Parkes\footnote{http://www.parkes.atnf.csiro.au/} radio telescope, using the new K-band receiver which operates from 16\,--26\,GHz, during three nights spanning 2008, September 17\,--\,19. During this period, 244 sources were observed simultaneously in both \nh\, (1,1) and (2,2). 

The K-band receiver, which was partially commissioned in September 2008, was used in conjunction with the digital filterbank-3 (DFB) backend which contains dual digitisers and Compact Array Broadband Backend (CABB)\footnote{ The CABB hardware was originally developed for the Australia Telescope Compact Array (ATCA) and was also used in the DFB3 backend at Parkes. See http://www.narrabri.atnf.csiro.au/\-observing/CABB.html} processors\footnote{see http://www.parkes.atnf.csiro.au/observing/documentation/\-software/CORREL/index.html}. Although the Parkes telescope is 64\,m in diameter, only the inner 55\,m is used for observations at 23\,GHz.

The \nh\, (1,1) and (2,2) observations were undertaken in position switch mode, using a bandwidth of 128\,MHz centred at 23708\,MHz to enable simultaneous observations of the two transitions.  Maser velocities \citep{pestalozzi05} were adopted as the rest velocity for each of the sources targeted. For those sources devoid of maser associations i.e. class R and MM, the velocity of the nearest coincident methanol maser was adopted as the rest velocity. We used 8192 channels to produce an expected velocity resolution of 0.2\,km/s.

The pointing accuracy of the Parkes telescope is typically 10\,--15\, arcsec which is smaller than the beamsize of 58 arcsec. The Tsys ranged from 85 to 132\,K during the period of the observations (12\,--15\,hr observing shifts), with a median of 96\,K. 

Typical integration times were 5 minutes per source. The rms noise in the spectra ranges from 0.13 to 0.30\,Jy/beam for the \nhone\, transition and 0.13 to 0.25\,Jy/beam for the (2,2) transition with the exception of one source (G\,294.989-1.720) which had an rms of 0.86 and 0.87\,Jy/beam for each of the transitions, respectively. The full set of spectra, including the sample spectra given in Figure \ref{fig:sample:spectra}, are presented in the Appendix.

\begin{figure*}
\begin{center}
\begin{tabular}{cc}
 \includegraphics[angle=-90,width=0.45\hsize]{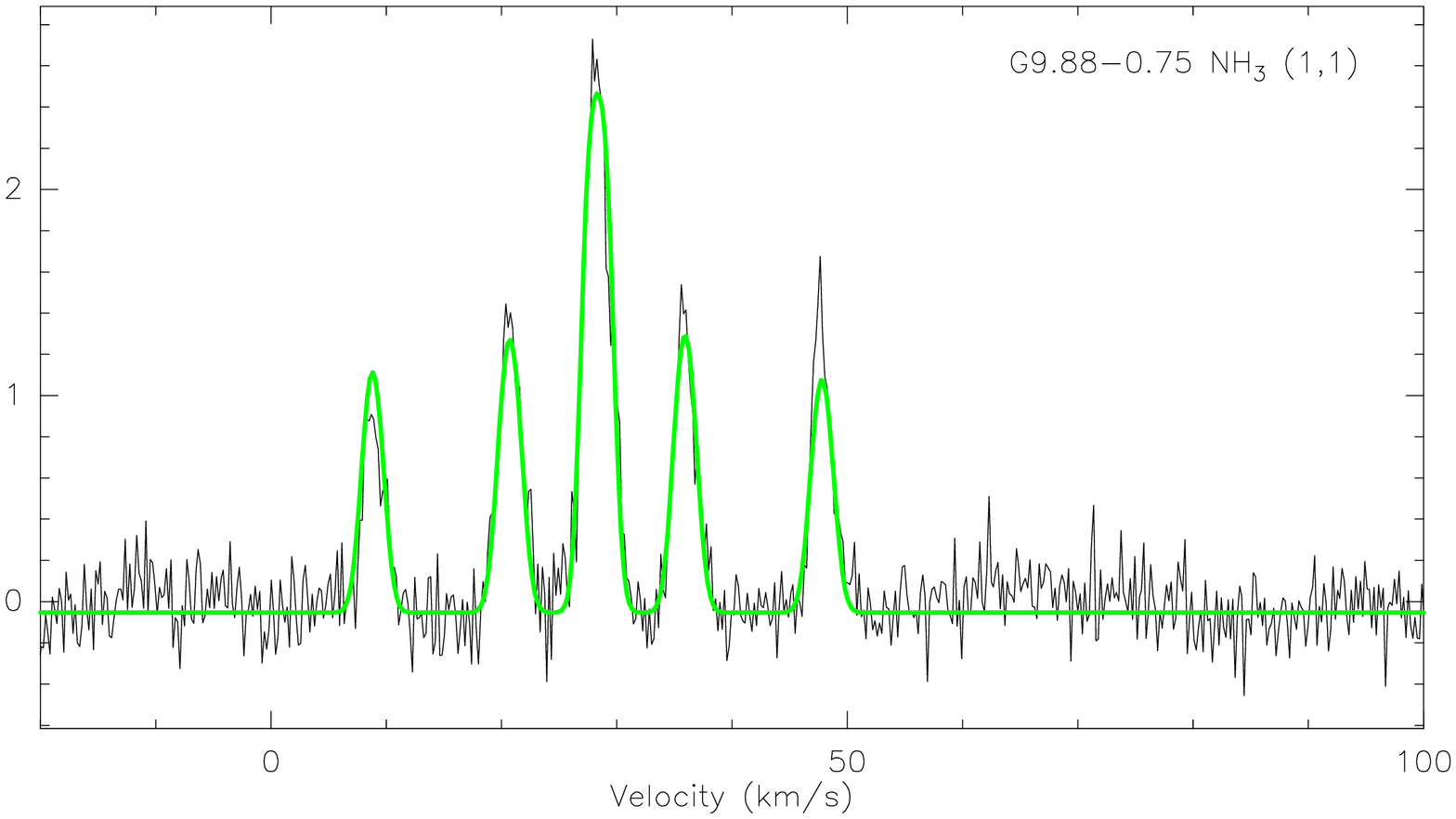} &
 \includegraphics[angle=-90,width=0.45\hsize]{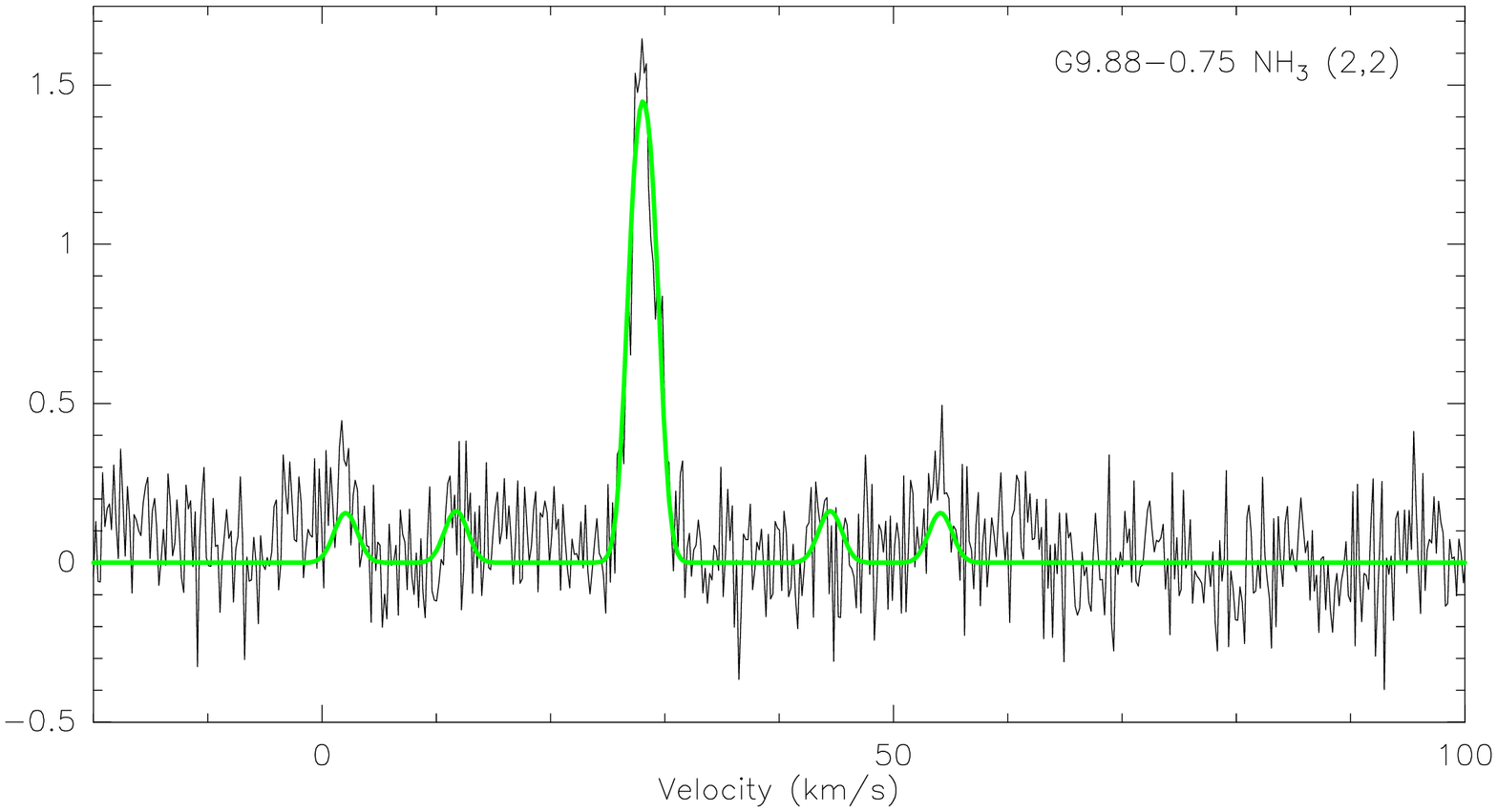} \\
 \includegraphics[angle=-90,width=0.45\hsize]{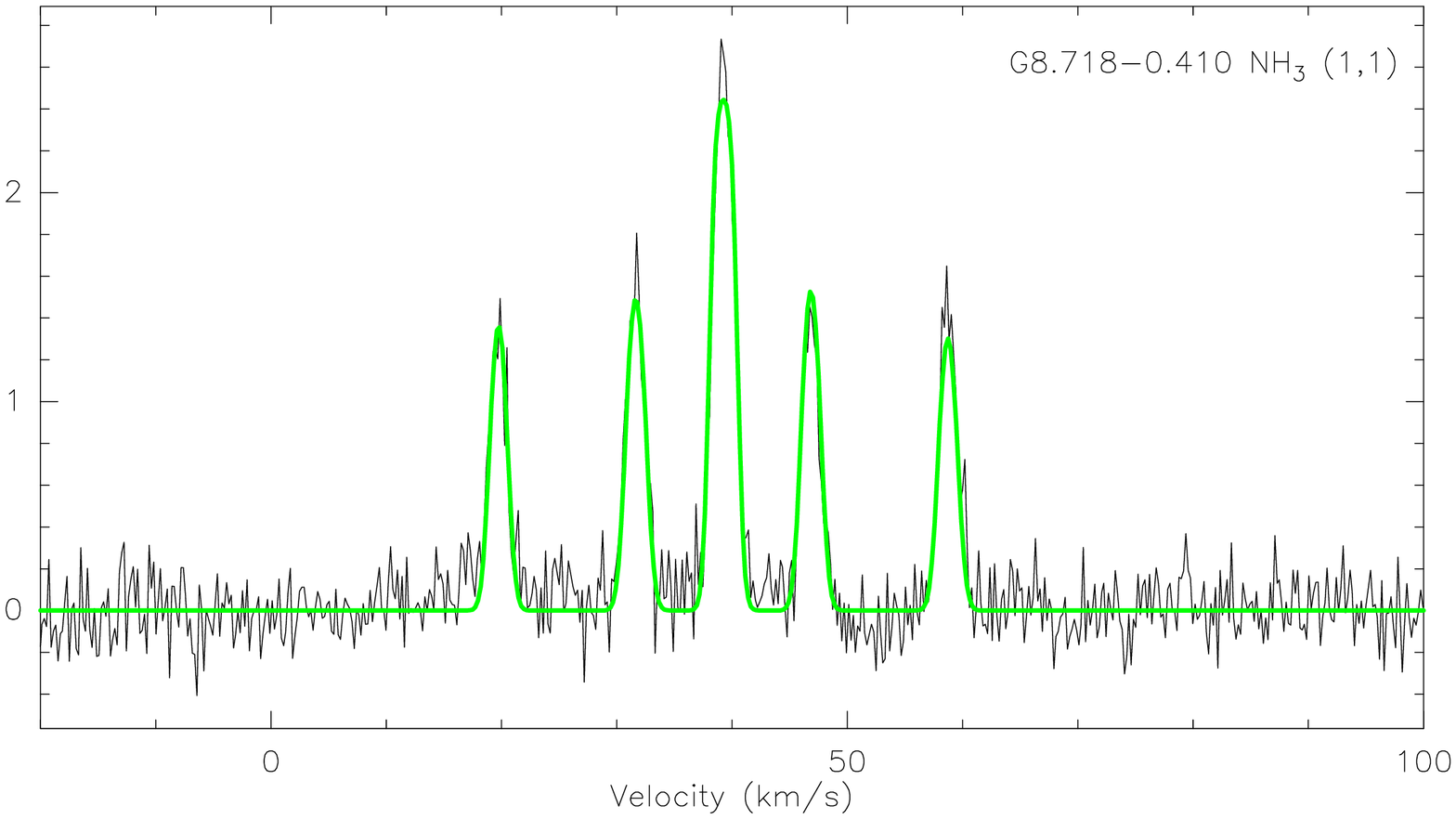} &
 \includegraphics[angle=-90,width=0.45\hsize]{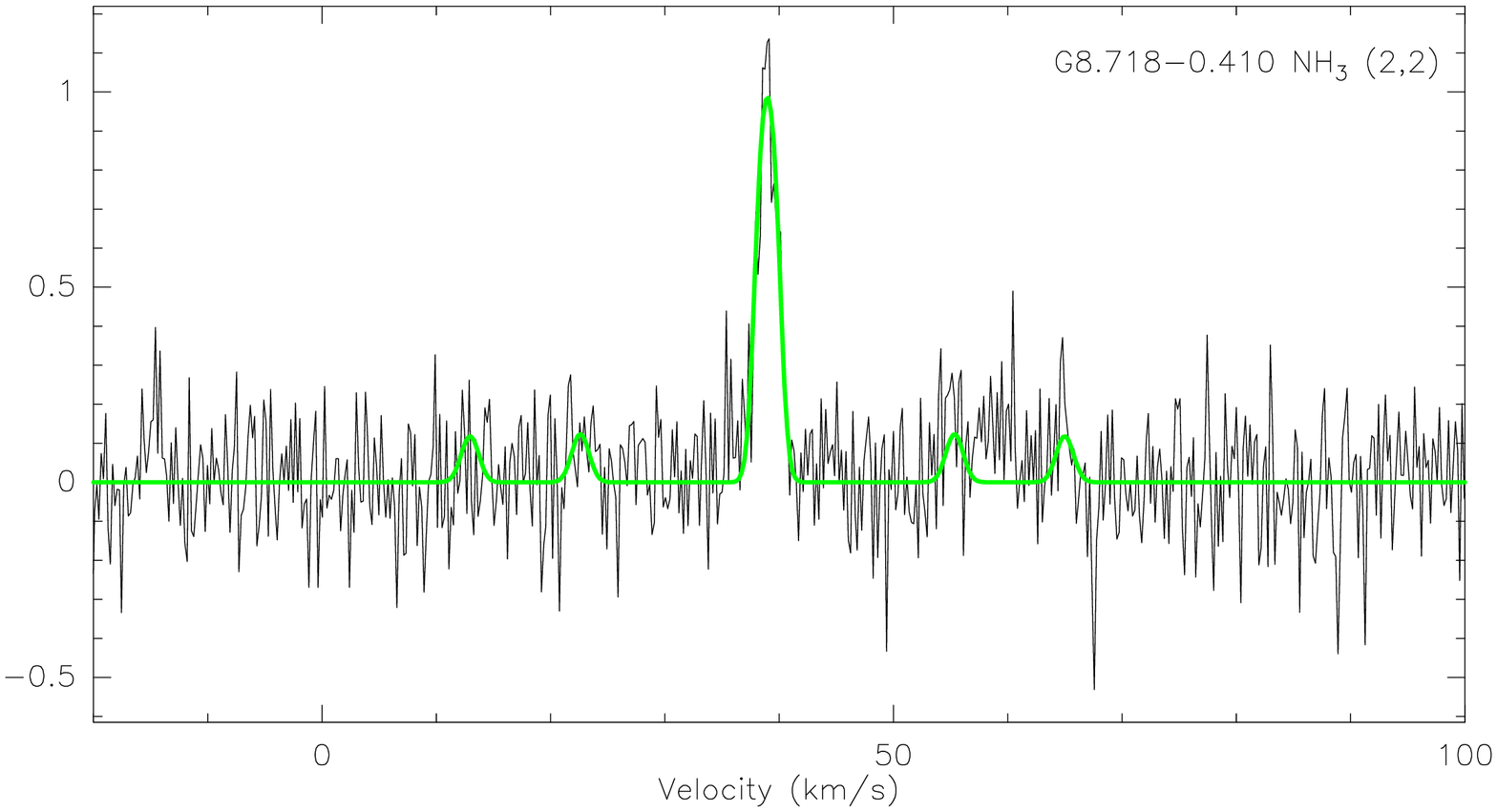} \\
\end{tabular}
\end{center}
\caption{Sample \nhone\, and (2,2) spectra, shown left to right, respectively. The y-axis is in units of Jy/beam. The green lines on the spectra indicate the fits as determined with CLASS.  Top Row: G9.88-0.75 a class M (see section~\ref{sec:obs}) source, and bottom: G8.718-0.410 a class MM source. The full set of \nhone\, and (2,2) spectra can be found in the Appendix. Whilst a sample of the full appendix is included here (Fig. \ref{fig:spectra}), the entire appendix of spectra can be found in the online version.}\label{fig:sample:spectra}
\end{figure*}

%__________________________________________________________________

%\vspace{-0.5cm}
\section{Data Processing}\label{sec:processing}

The data were reduced using the ATNF Spectral Analysis Package (ASAP)\footnote{http://www.atnf.csiro.au/computing/software/asap/} and fit using the Continuum and Line Analysis Single-dish Software (CLASS)\footnote{http://www.iram.fr/IRAMFR/GILDAS/doc/html/class-html/class.html} package, using standard procedures as described in the respective manuals.

\subsection{Data Reduction}

ASAP is a new software package developed by the ATNF to reduce single-dish single-pointing spectral lines observations, with particular application to the ATNF suite of telescopes.

During the reduction procedure, the data were read into ASAP, an auto\_quotient\footnote{The ASAP {\it auto\_quotient} command combines the nearest source and reference spectra, dividing and subtracting the latter off.} was applied, as well as a first order polynomial before the data were frequency-aligned. The data were then averaged over time and polarisation, the gain elevation correction was applied, as was an opacity correction of 10 per cent (priv. communication J. Reynolds). The rest frequency was then set (23.694512 and 23.7226336\,GHz for the \nhone\, and \nhtwo\, transitions, respectively) and a third order polynomial was applied.

Comparison of the main hyperfine components \vlsr\, of the \nhone\, and \nhtwo\, transitions reveals a small velocity shift between the (1,1) and (2,2) spectra. The mean offset in velocity between the two transitions is 0.24\,km/s with a standard deviation of 0.26\,km/s.  As mentioned in section~\ref{sec:obs}, the expected resolution of these ammonia observations is 0.2\,km/s. The rest frequencies, which were obtained from the Lovas catalogue, are believed to be less accurate than the velocity shift seen in our spectra.

\subsection{\nh\, Fitting}\label{sec:fitting}

Each of the \nhone\, and \nhtwo\, spectra were fit using the standard \nhone\, and \nhtwo\, methods, respectively, in CLASS.  Under the assumption that each of the hyperfine components have equal excitation temperature and the same Gaussian velocity structure, these methods return the calculated flux density, rest velocity (\vlsr), linewidth (\delv), \& optical depth ($\tau$) as well as the corresponding uncertainties to the fits for each source. A 3-$\sigma$ cutoff was used to define non-detections for both transitions. 

While the data quality is mostly of a high standard, a subset of the spectra show approximately sinusoidal variations in the baseline,  typically with period 40$-$100\kms. Attempting to use polynomial fitting to remove baseline variations with a similar velocity scale as the hyperfine structure ($\sim$50\kms) may substantially alter the structure of the spectra. While the line $\vlsr$ and $\delv$ are likely robust (the line width of the main component and satellites are typically much smaller than the baseline variations), the ratios of measured flux density, both for the hyperfines within a transition and between different transitions, are not. Derived optical depths and temperatures should therefore be treated with caution for these sources. Any spectra showing this anomaly are flagged as BR (baseline ripple) in column 11 of Table~\ref{tab:nh3_11_22_fits}. Some sources, particularly those towards the Galactic centre, show multiple blended velocity components, making the fits less reliable. These data are flagged as BL (Blended) in column 11 of Table~\ref{tab:nh3_11_22_fits}. All other spectra are considered reliable and marked as an R in this column.

The parameters derived from the fits to both the \nhone\, and \nhtwo\, spectra are given in Table \ref{tab:nh3_11_22_fits}.  The source name is given in column 1, in right ascension order, using names consistent with \citet{hill05}. The source class is given in column 2. Columns 3, 4, 5 and 6 present the flux density, \vlsr, linewidth (\delv) and the optical depth for the \nhone\, transition. Columns 7\,-\,10 present these parameters for the \nhtwo\, transition. For sources with an \nhone\, detection but no \nhtwo\, detection, a 3-$\sigma$ upper limit is derived from the \nhtwo\, spectra which is given in column 7, and no values are listed in columns 8\,--\,10.

%\fontsize{7.5}{9}\selectfont \medium
\begin {table*}
\caption{$\nhone$ and (2,2) fit results.} %Parameters extracted from the fits to the
\label{tab:nh3_11_22_fits}
\scriptsize
\begin{tabular}{@{}l|@{}l|@{}l|@{}c|@{}c@{}|@{}c|@{}c|@{}c@{}|@{}c@{}|@{}c|@{}c|@{}c|@{}c|@{}c@{}|} \hline 
Right & Declin- &Source & $^\alpha$ & Flux Den.$^{1,1}$ & V$_{\rm LSR}^{1,1}$ & $\Delta$V$^{1,1}$ & $\tau_{\rm main}^{1,1}$ & Flux Den.$^{2,2}$ & V$_{\rm LSR}^{2,2}$ & $\Delta$V$^{2,2}$ & $\tau_{\rm main}^{2,2}$ & Base- \\
Ascension & ation&       & & (Jy/beam)     & (km/s)      & (km/s)    &    &  (Jy/beam)     & (km/s) & (km/s)       & &Line$^\beta$  \\ \hline
09 03 33.1 & -48 28 08 &  G269.15-1.13 & M &1.65$\pm$0.16  & 10.50$\pm$0.06  &  3.20$\pm$0.17  & 0.32$\pm$0.25 & 0.96$\pm$0.06  & 10.40$\pm$0.11  &  3.63$\pm$0.27  & 0.10$\pm$0.49 & BR   \\
11 11 33.9 &-61 21 22 & G291.256-0.769 & MM & 3.81$\pm$0.19  & -24.00$\pm$0.02  &  1.79$\pm$0.06  & 1.07$\pm$0.15 & 1.34$\pm$0.09  & -24.20$\pm$0.04  &  2.13$\pm$0.13  & 0.10$\pm$2.47 & R \\
11 11 39.4& -61 19 46 & G291.256-0.743 & MM & 2.37$\pm$0.13  & -24.20$\pm$0.03  &  2.95$\pm$0.09  & 0.57$\pm$0.15 & 1.32$\pm$0.05  & -24.40$\pm$0.06  &  2.93$\pm$0.15  & 0.10$\pm$0.11 & R  \\
11 12 15.0 & -61 17 30 & G291.309-0.681 & MM & 1.28$\pm$0.13  & -24.80$\pm$0.06  &  2.84$\pm$0.16  & 0.40$\pm$0.27 & 0.64$\pm$0.06  & -25.10$\pm$0.12  &  2.95$\pm$0.25  & 0.10$\pm$0.71 & R \\
11 14 57.8 & -61 11 40& G291.576-0.468 & MM & 0.84$\pm$0.21  & 13.40$\pm$0.10  &  1.91$\pm$0.25  & 1.00$\pm$0.74 & $<$0.63  & - & - & - & BR \\
11 15 05.9 &-61 09 46 & G291.58-0.53 & M &0.80$\pm$0.09  & 14.20$\pm$0.14  &  5.26$\pm$0.29  & 0.18$\pm$0.24 & $<$0.33  & - & - & - & R\\
11 39 13.8 & -63 29 10 & G294.97-1.7 & R &1.83$\pm$0.19  & -8.29$\pm$0.04  &  1.71$\pm$0.10  & 0.61$\pm$0.28 & $<$0.55  & - & - & - &BR \\
13 08 13.5 & -62 10 20 & G304.890+0.636 & MM & 0.73$\pm$0.23  & -36.10$\pm$0.09  &  1.56$\pm$0.34  & 0.62$\pm$0.69 & 0.12$\pm$0.03  & -34.70$\pm$1.21  &  10.80$\pm$2.57  & 0.10$\pm$1.08 &R\\  
13 10 40.5 & -62 34 53 & G305.145+0.208 & MM & 0.98$\pm$0.30  & -17.70$\pm$0.09  &  1.15$\pm$0.23  & 1.71$\pm$1.09 & $<$0.39  & - & - & - &BR\\
13 10 43.3 & -62 43 05  & G305.137+0.069 & MM & 2.24$\pm$0.21  & -36.30$\pm$0.05  &  2.25$\pm$0.12  & 1.51$\pm$0.32 & $<$0.42  & - & - & - &R\\
13 11 14.7 & -62 47 21& G305.192-0.006 & M &2.12$\pm$0.15  & -33.30$\pm$0.07  &  4.26$\pm$0.16  & 0.64$\pm$0.19 & 0.90$\pm$0.06  & -33.50$\pm$0.15  &  4.77$\pm$0.41  & 0.10$\pm$1.40 &BR  \\
13 11 14.1 & -62 34 45& G305.21+0.21 & M &1.67$\pm$0.17  & -41.70$\pm$0.12  &  6.97$\pm$0.34  & 0.26$\pm$0.20 & 1.21$\pm$0.04  & -42.00$\pm$0.14  &  8.58$\pm$0.34  & 0.10$\pm$0.10  &BR/BL \\
13 11 15.8 & -62 46 41&G305.197+0.007 & MM & 3.50$\pm$0.22  & -32.80$\pm$0.04  &  2.54$\pm$0.10  & 2.01$\pm$0.24 & 1.16$\pm$0.36  & -33.20$\pm$0.09  &  3.00$\pm$0.34  & 0.48$\pm$0.72  & BR  \\
13 11 19.9 & -62 30 29& G305.226+0.275 & MM & 5.16$\pm$0.17  & -39.30$\pm$0.02  &  2.95$\pm$0.05  & 1.36$\pm$0.10 & 2.98$\pm$0.40  & -39.60$\pm$0.05  &  2.68$\pm$0.13  & 1.42$\pm$0.38  &R\\
13 11 21.0 & -62 29 49& G305.228+0.286 & MM & 3.98$\pm$0.16  & -39.60$\pm$0.03  &  2.94$\pm$0.07  & 1.50$\pm$0.14 & 2.85$\pm$0.48  & -40.00$\pm$0.08  &  2.55$\pm$0.16  & 2.66$\pm$0.64 &R \\
13 11 26.6 & -62 31 17 & G305.238+0.261 & MM & 2.83$\pm$0.19  & -38.70$\pm$0.04  &  2.46$\pm$0.10  & 1.18$\pm$0.22 & 1.18$\pm$0.35  & -38.80$\pm$0.08  &  2.60$\pm$0.28  & 0.42$\pm$0.67  &R\\
13 11 32.6 & -62 32 13 & G305.248+0.245 & M &1.19$\pm$0.20  & -37.10$\pm$0.11  &  3.01$\pm$0.35  & 1.03$\pm$0.48 & $<$0.27  & - & - & -&BR \\
13 11 35.7 & -62 48 17& G305.233-0.023 & MM & 4.33$\pm$0.22  & -28.80$\pm$0.03  &  2.56$\pm$0.07  & 1.86$\pm$0.19 & 1.61$\pm$0.44  & -28.90$\pm$0.08  &  2.61$\pm$0.24  & 0.99$\pm$0.69 &R \\
13 11 54.3 & -62 47 20 & G305.269-0.010 & MM & 4.12$\pm$0.18  & -32.20$\pm$0.03  &  3.23$\pm$0.08  & 1.58$\pm$0.15 & 1.52$\pm$0.06  & -32.10$\pm$0.07  &  3.81$\pm$0.17  & 0.10$\pm$0.16 &R \\
13 12 30.5 & -62 34 43 & G305.355+0.194 & MM & 3.30$\pm$0.29  & -37.70$\pm$0.05  &  2.21$\pm$0.13  & 1.98$\pm$0.33 & 1.85$\pm$0.53  & -38.00$\pm$0.08  &  2.62$\pm$0.30  & 1.17$\pm$0.77 &R \\
13 12 31.6 & -62 34 11& G305.37+0.21 & R &0.85$\pm$0.16  & -37.40$\pm$0.30  &  6.69$\pm$0.67  & 0.79$\pm$0.45 & $<$0.28  & - & - & - &BR/BL \\
13 12 33.9 & -62 35 15& G305.362+0.185 & M &4.43$\pm$0.22  & -37.50$\pm$0.04  &  2.96$\pm$0.09  & 1.51$\pm$0.17 & 2.94$\pm$0.60  & -37.70$\pm$0.05  &  2.68$\pm$0.21  & 1.07$\pm$0.53  &R\\
13 12 35.1 & -62 37 15& G305.361+0.151 & M &0.80$\pm$0.07  & -38.20$\pm$0.12  &  3.20$\pm$0.38  & 0.10$\pm$0.47 & $<$0.39  & - & - & - &R \\
13 12 36.3 & -62 33 39 & G305.368+0.211$^\dagger$ & R &1.26$\pm$0.16  & -35.00$\pm$0.16  &  4.76$\pm$0.31  & 1.01$\pm$0.38 & 2.99$\pm$0.54  & -35.70$\pm$0.19  &  4.00$\pm$0.26  & 6.12$\pm$1.32 &BR \\
13 12 37.4 & -62 36 35& G305.340-0.172$^\dagger$ & MM & 1.56$\pm$0.22  & -39.20$\pm$0.10  &  3.02$\pm$0.25  & 1.19$\pm$0.44 & 2.01$\pm$0.61  & -39.50$\pm$0.17  &  2.69$\pm$0.32  & 3.40$\pm$1.32 &R \\
13 13 58.7 & -62 25 05 & G305.538+0.340 & MM & 1.27$\pm$0.13  & -34.70$\pm$0.09  &  3.25$\pm$0.19  & 1.28$\pm$0.35 & $<$0.38  & - & - & - &R\\
13 14 21.2 & -62 44 33 & G305.55+0.01 & M &0.87$\pm$0.15  & -39.30$\pm$0.10  &  2.65$\pm$0.25  & 0.80$\pm$0.50 & 0.41$\pm$0.08  & -39.30$\pm$0.20  &  3.95$\pm$1.41  & 0.10$\pm$0.13 &R \\
13 14 22.3 & -62 46 09 & G305.552+0.012 & MM & 0.48$\pm$0.04  & -36.50$\pm$0.16  &  3.66$\pm$0.30  & 0.10$\pm$0.87 & $<$0.23  & - & - & - &R\\
13 14 27.0 & -62 44 33& G305.561+0.012 & R &0.94$\pm$0.14  & -39.40$\pm$0.07  &  2.81$\pm$0.21  & 0.24$\pm$0.33 & 0.57$\pm$0.05  & -39.80$\pm$0.13  &  2.96$\pm$0.28  & 0.10$\pm$0.29 &R \\
13 16 31.4 & -62 59 01 &G305.776-0.251 & MM & 1.67$\pm$0.28  & -29.80$\pm$0.05  &  1.41$\pm$0.15  & 0.87$\pm$0.48 & 0.65$\pm$0.12  & -30.10$\pm$0.10  &  1.36$\pm$0.28  & 0.10$\pm$20.00 &R \\ 
13 16 43.2 & -62 58 37 & G305.81-0.25 & MR &1.11$\pm$0.12  & -31.40$\pm$0.10  &  4.09$\pm$0.22  & 0.84$\pm$0.31 & 0.55$\pm$0.04  & -31.80$\pm$0.18  &  5.42$\pm$0.43  & 0.10$\pm$0.35 &R \\
13 16 58.4 & -62 55 25 & G305.833-0.196 & MM & 1.21$\pm$0.17  & -33.20$\pm$0.07  &  2.31$\pm$0.20  & 1.00$\pm$0.41 & 0.31$\pm$0.08  & -33.80$\pm$0.18  &  1.35$\pm$0.44  & 0.10$\pm$1.66 &R \\
13 21 21.7 & -63 00 35 & G306.33-0.3 & M &0.93$\pm$0.16  & -19.20$\pm$0.10  &  2.34$\pm$0.23  & 1.20$\pm$0.56 & 0.17$\pm$0.04  & -19.10$\pm$1.06  &  9.17$\pm$3.01  & 0.10$\pm$0.44 &R \\
13 21 32.3 & -62 58 35 & G306.343-0.302 & MM & 1.02$\pm$0.24  & -19.10$\pm$0.11  &  1.92$\pm$0.28  & 2.08$\pm$0.90 & $<$0.99  & - & - & - &R \\
13 21 34.7 & -63 00 03&G306.345-0.345$^\dagger$ & MM & 0.29$\pm$0.04  & -19.10$\pm$0.25  &  3.56$\pm$0.58  & 0.10$\pm$0.25 & $<$0.71  & - & - & - &R \\ 
13 50 38.2 & -61 34 28 &G309.917+0.494 & MM & 0.45$\pm$0.14  & -57.00$\pm$0.26  &  3.43$\pm$0.48  & 1.26$\pm$1.01 & $<$0.13  & - & - & - &R\\
13 50 41.6 & -61 35 16& G309.92+0.4 & M &0.92$\pm$0.14  & -57.50$\pm$0.09  &  2.71$\pm$0.20  & 0.38$\pm$0.37 & $<$0.49  & - & - & - &BR\\
15 00 54.3 & -58 58 53 & G318.92-0.68 & M &3.87$\pm$0.20  & -34.30$\pm$0.03  &  2.26$\pm$0.07  & 1.60$\pm$0.17 & 2.12$\pm$0.48  & -34.40$\pm$0.07  &  2.23$\pm$0.21  & 1.82$\pm$0.70 &R \\
15 31 44.5 & -56 30 51 &G323.74-0.3 & M &1.85$\pm$0.17  & -49.60$\pm$0.05  &  2.53$\pm$0.14  & 1.00$\pm$0.27 & 1.26$\pm$0.39  & -49.70$\pm$0.09  &  2.39$\pm$0.29  & 1.27$\pm$0.85 &R \\
16 11 26.9 & -51 41 57& G331.279-0.189$^\dagger$ & M &2.38$\pm$0.16  & -87.80$\pm$0.05  &  3.19$\pm$0.12  & 1.47$\pm$0.23 & 3.05$\pm$0.48  & -88.20$\pm$0.11  &  2.98$\pm$0.21  & 4.28$\pm$0.84 &R \\
16 19 37.6 & -51 03 16 & G332.648-0.606$^\dagger$ & M &1.36$\pm$0.15  & -49.50$\pm$0.10  &  3.78$\pm$0.22  & 1.18$\pm$0.35 & 1.55$\pm$0.47  & -49.80$\pm$0.18  &  2.91$\pm$0.38  & 3.03$\pm$1.24 &R \\
16 19 48.6 & -51 02 12& G332.646-0.647A & MM & 1.20$\pm$0.01  & -57.70$\pm$0.08  &  3.91$\pm$0.14  & 0.39$\pm$0.10 & 0.54$\pm$0.06  & -57.90$\pm$0.13  &  2.17$\pm$0.28  & 0.10$\pm$0.07  &R\\
16 19 48.6 & -51 02 12& G332.646-0.647B & MM & 0.80$\pm$0.01  & -47.80$\pm$0.09  &  3.93$\pm$0.17  & 0.10$\pm$0.03 & 0.49$\pm$0.07  & -48.20$\pm$0.13  &  1.94$\pm$0.30  & 0.10$\pm$0.08  &R\\
16 19 51.7 & -51 01 28& G332.695-0.609 & MM & 2.83$\pm$0.23  & -47.80$\pm$0.05  &  2.51$\pm$0.12  & 2.34$\pm$0.34 & $<$0.53  & - & - & - &R \\
16 20 02.7 & -51 00 32& G332.725-0.62 & M &3.99$\pm$0.36  & -49.90$\pm$0.03  &  1.21$\pm$0.07  & 2.64$\pm$0.40 & $<$0.63  & - & - & - &R\\
16 20 06.9 & -51 00 00& G332.627-0.511 & MM & 2.20$\pm$0.27  & -50.00$\pm$0.04  &  1.45$\pm$0.10  & 1.57$\pm$0.42 & 0.68$\pm$0.08  & -50.30$\pm$0.14  &  2.05$\pm$0.35  & 0.58$\pm$0.54 &R \\
16 20 12.0 & -50 53 20 & G332.827-0.552 & MM & 1.50$\pm$0.13  & -56.30$\pm$0.12  &  5.03$\pm$0.22  & 1.18$\pm$0.27 & 0.63$\pm$0.04  & -56.10$\pm$0.20  &  6.74$\pm$0.41  & 0.10$\pm$0.96 &R/BL \\
17 46 58.8 & -28 45 12& G0.331-0.164 & MM & 1.71$\pm$0.13  & 20.40$\pm$0.07  &  3.49$\pm$0.18  & 1.28$\pm$0.25 & 0.34$\pm$0.04  &-&-&-& BR/BL\\%& 21.50$\pm$0.43  &  8.58$\pm$1.55  & 0.10$\pm$0.10 &BR  \\
17 47 01.2 & -28 45 36 & G0.310-0.170 & MM & 0.85$\pm$0.11  & 19.70$\pm$0.18  &  4.96$\pm$0.38  & 0.98$\pm$0.41 & 0.24$\pm$0.06& -&-&-&BR/BL \\%  & 20.80$\pm$0.59  &  9.22$\pm$1.84  & 0.10$\pm$21.10 &BR/BL  \\
17 47 09.7 & -28 46 08 & G0.32-0.20 & MR &4.33$\pm$0.24  & 18.80$\pm$0.02  &  1.73$\pm$0.06  & 1.77$\pm$0.19 & 2.30$\pm$0.52  & 18.50$\pm$0.06  &  1.53$\pm$0.13  & 2.05$\pm$0.75 &BR \\
17 48 36.4 & -28 02 31& G1.105-0.098 & MM & 2.56$\pm$0.01  & -16.80$\pm$0.03  &  2.65$\pm$0.07  & 0.11$\pm$0.03 & 1.60$\pm$0.07  & -17.00$\pm$0.06  &  2.17$\pm$0.14  & 0.51$\pm$0.14 &BR/BL \\
17 48 42.5 & -28 01 35 & G1.13-0.11 & R &2.56$\pm$0.01  & -15.20$\pm$0.03  &  2.93$\pm$0.06  & 0.10$\pm$0.02 & 1.60$\pm$0.03  & -15.40$\pm$0.05  &  2.75$\pm$0.10  & 0.12$\pm$0.06  &BR/BL \\
17 50 18.8 & -28 53 19& G0.549-0.868 & MM & 3.72$\pm$0.28  & 17.30$\pm$0.02  &  1.35$\pm$0.07  & 1.13$\pm$0.23 & 1.13$\pm$0.10  & 17.10$\pm$0.06  &  1.66$\pm$0.18  & 0.10$\pm$0.20 &BR   \\
17 50 25.5 & -28 50 15 & G0.627-0.848 & MM & 5.37$\pm$0.42  & 18.30$\pm$0.03  &  1.88$\pm$0.09  & 1.73$\pm$0.28 & $<$1.02  & - & - & - &BR \\
17 50 26.1 & -28 52 31& G0.600-0.871 & MM & 4.42$\pm$0.24  & 18.10$\pm$0.02  &  1.59$\pm$0.05  & 1.55$\pm$0.19 & 1.13$\pm$0.25  & 18.00$\pm$0.06  &  1.86$\pm$0.20  & 0.10$\pm$5.31 &R  \\
17 50 46.5 & -26 39 44 & G2.54+0.20 & M &3.90$\pm$0.33  & 9.96$\pm$0.03  &  1.48$\pm$0.08  & 2.42$\pm$0.35 & 0.89$\pm$0.07  & 9.78$\pm$0.08  &  2.33$\pm$0.24  & 0.10$\pm$1.30 &BR \\
17 59 07.5 & -24 19 19 &G5.504-0.246 & MM & 2.34$\pm$0.26  & 21.70$\pm$0.05  &  1.69$\pm$0.11  & 2.18$\pm$0.46 & 0.50$\pm$0.06  & 21.20$\pm$0.14  &  2.16$\pm$0.29  & 0.10$\pm$0.62 &R \\
18 00 30.4 & -24 03 59& G5.89-0.39$^\dagger$ & R &2.36$\pm$0.13  & 9.52$\pm$0.05  &  3.78$\pm$0.10  & 0.57$\pm$0.15 & 2.51$\pm$0.34  & 9.40$\pm$0.07  &  3.72$\pm$0.18  & 1.37$\pm$0.38  &R\\
18 00 40.9 & -24 04 12& G5.90-0.42 & M &3.57$\pm$0.18  & 7.63$\pm$0.04  &  3.36$\pm$0.09  & 1.67$\pm$0.18 & 1.54$\pm$0.40  & 7.42$\pm$0.07  &  3.67$\pm$0.34  & 0.30$\pm$0.56  &R\\
18 00 43.8 & -24 04 52& G5.90-0.44 & MM & 3.78$\pm$0.21  & 9.81$\pm$0.03  &  1.99$\pm$0.06  & 1.29$\pm$0.18 & 2.31$\pm$0.53  & 9.54$\pm$0.04  &  2.06$\pm$0.18  & 0.86$\pm$0.56 &R \\
18 02 49.3 & -21 48 34 &G8.111+0.257 & MM & 0.97$\pm$0.07  & 19.50$\pm$0.07  &  1.99$\pm$0.18  & 0.10$\pm$0.22 & 0.53$\pm$0.08  & 19.00$\pm$0.15  &  2.03$\pm$0.29  & 0.10$\pm$0.35 &R \\
18 02 52.8 & -21 47 54& G8.127+0.255 & MM & 1.75$\pm$0.19  & 19.80$\pm$0.06  &  2.64$\pm$0.16  & 0.81$\pm$0.30 & 0.57$\pm$0.08  & 19.60$\pm$0.23  &  3.10$\pm$0.71  & 0.10$\pm$0.16 &BL \\
18 02 56.2 & -21 47 38&G8.138+0.246 & MM & 1.65$\pm$0.16  & 19.20$\pm$0.07  &  3.35$\pm$0.15  & 0.67$\pm$0.27 & 0.57$\pm$0.06  & 19.10$\pm$0.17  &  3.39$\pm$0.34  & 0.10$\pm$0.38 &R \\
18 03 02.0 & -21 48 02 & G8.13+0.22 & MR &4.46$\pm$0.28  & 19.40$\pm$0.03  &  2.05$\pm$0.08  & 1.37$\pm$0.20 & 1.55$\pm$0.09  & 19.20$\pm$0.06  &  2.15$\pm$0.17  & 0.10$\pm$0.25 &R \\ 
18 03 26.9 & -24 22 29& G5.948-1.125 & MM & 1.22$\pm$0.40  & 9.49$\pm$0.08  &  1.06$\pm$0.26  & 0.92$\pm$0.91 & $<$0.50  & - & - & - &R \\
18 03 29.2 & -24 21 49& G5.962-1.128 & MM & 2.73$\pm$0.45  & 9.28$\pm$0.04  &  1.13$\pm$0.13  & 1.44$\pm$0.55 & 0.58$\pm$0.13  & 9.01$\pm$0.15  &  1.32$\pm$0.32  & 0.10$\pm$6.74 &BR \\
18 03 33.9 & -24 21 41& G5.975-1.146 & MM & 2.69$\pm$0.36  & 8.59$\pm$0.04  &  1.31$\pm$0.11  & 1.49$\pm$0.46 & 1.17$\pm$0.15  & 8.38$\pm$0.11  &  1.54$\pm$0.23  & 1.14$\pm$0.05 &R \\
18 03 36.8 & -24 22 08& G5.971-1.158 & MM & 1.78$\pm$0.34  & 8.31$\pm$0.07  &  1.59$\pm$0.20  & 1.15$\pm$0.56 & $<$0.31  & - & - & - &R\\ 
18 06 14.8 & -20 31 29&G9.63+0.19 & MR &3.75$\pm$0.17  & 4.34$\pm$0.04  &  3.63$\pm$0.09  & 1.39$\pm$0.15 & 2.35$\pm$0.42  & 4.29$\pm$0.08  &  4.11$\pm$0.31  & 1.26$\pm$0.49 &R \\
18 06 18.9 & -21 37 21& G8.68-0.36 & M &8.90$\pm$0.19  & 35.60$\pm$0.02  &  3.47$\pm$0.04  & 2.50$\pm$0.10 & 2.56$\pm$0.36  & 35.30$\pm$0.05  &  4.06$\pm$0.20  & 0.22$\pm$0.31 &R \\
18 06 23.5 & -21 36 57& G8.686-0.366 & M &5.87$\pm$0.19  & 37.30$\pm$0.04  &  4.29$\pm$0.07  & 2.18$\pm$0.13 & 2.41$\pm$0.36  & 37.30$\pm$0.07  &  4.27$\pm$0.21  & 1.12$\pm$0.39 &R \\
18 06 24.6 & -21 40 01& G8.644-0.395 & MM & 4.68$\pm$0.37  & 39.20$\pm$0.03  &  1.20$\pm$0.06  & 2.69$\pm$0.36 & 0.80$\pm$0.11  & 39.10$\pm$0.09  &  1.69$\pm$0.21  & 0.10$\pm$1.15 &R \\
18 06 26.4 & -21 35 29& G8.713-0.364 & MM & 6.27$\pm$0.37  & 38.00$\pm$0.03  &  2.06$\pm$0.07  & 3.75$\pm$0.33 & 0.89$\pm$0.08  & 38.00$\pm$0.09  &  2.20$\pm$0.25  & 0.10$\pm$0.46 &R \\
18 06 28.7 & -21 34 17& G8.735-0.362 & MM & 4.99$\pm$0.21  & 39.40$\pm$0.03  &  2.41$\pm$0.06  & 2.00$\pm$0.15 & 1.27$\pm$0.42  & 39.40$\pm$0.07  &  2.39$\pm$0.28  & 0.42$\pm$0.75 &R \\
18 06 36.1 & -21 36 01& G8.724-0.401 & MM & 8.61$\pm$0.35  & 39.30$\pm$0.02  &  1.63$\pm$0.04  & 3.29$\pm$0.21 & 1.12$\pm$0.06  & 39.00$\pm$0.06  &  2.36$\pm$0.18  & 0.10$\pm$1.77 &R \\
18 06 36.7 & -21 37 05& G8.709-0.412 & MM & 5.54$\pm$0.22  & 39.40$\pm$0.02  &  2.05$\pm$0.05  & 2.01$\pm$0.15 & 0.99$\pm$0.07  & 39.10$\pm$0.07  &  2.33$\pm$0.19  & 0.10$\pm$0.94 &R \\
18 06 36.7 & -21 37 05& G8.718-0.410 & MM & 8.85$\pm$0.36  & 39.30$\pm$0.01  &  1.54$\pm$0.03  & 3.49$\pm$0.22 & 1.98$\pm$0.46  & 39.00$\pm$0.06  &  1.75$\pm$0.16  & 1.58$\pm$0.69 &R \\
18 08 38.5 & -19 51 48& G10.47+0.02$^\dagger$ & MR &4.19$\pm$0.17  & 67.10$\pm$0.08  &  6.31$\pm$0.15  & 1.95$\pm$0.14 & 6.38$\pm$0.47  & 67.10$\pm$0.11  &  5.53$\pm$0.16  & 5.78$\pm$0.51 &R/BL\\
18 08 45.9 & -20 05 42& G10.287-0.110 & MM & 3.71$\pm$0.22  & 14.10$\pm$0.05  &  3.10$\pm$0.11  & 1.87$\pm$0.22 & 2.15$\pm$0.58  & 13.90$\pm$0.10  &  2.21$\pm$0.22  & 2.52$\pm$0.98 &BR \\
18 08 49.3 & -20 05 58& G10.284-0.126 & M &4.18$\pm$0.20  & 14.00$\pm$0.03  &  2.79$\pm$0.08  & 1.74$\pm$0.17 & 2.41$\pm$0.47  & 13.70$\pm$0.08  &  2.78$\pm$0.21  & 1.87$\pm$0.63  &BR\\ 
\end{tabular}
\end{table*}
%\newpage 
\addtocounter{table}{-1} 
\begin {table*}
\caption{Cont.}
\scriptsize
 \begin{tabular}{@{}l|@{}l|@{}l@{}|@{}c@{}|@{}c@{}|@{}c|@{}c|@{}c@{}|@{}c@{}|@{}c|@{}c|@{}c|@{}c|@{}c@{}|} \hline 
Right & Declin- &Source & Class$^\alpha$ & Flux Den.$^{1,1}$ & V$_{\rm LSR}^{1,1}$ & $\Delta$V$^{1,1}$ & $\tau_{\rm main}^{1,1}$ & Flux Den.$^{2,2}$ & V$_{\rm LSR}^{2,2}$ & $\Delta$V$^{2,2}$ & $\tau_{\rm main}^{2,2}$ & Base-\\
Ascension & ation&       & & (Jy/beam)     & (km/s)      & (km/s)    &    &  (Jy/beam)     & (km/s) & (km/s)       & &Line$^\beta$   \\ \hline
18 08 52.7 & -20 05 58& G10.288-0.127 & MM & 2.38$\pm$0.17  & 14.20$\pm$0.05  &  3.10$\pm$0.13  & 0.94$\pm$0.21 & 0.96$\pm$0.05  & 13.80$\pm$0.11  &  4.11$\pm$0.24  & 0.10$\pm$0.18 &BR \\
18 08 56.1 & -20 05 50& G10.29-0.14 & MR &2.11$\pm$0.15  & 13.50$\pm$0.07  &  4.16$\pm$0.15  & 1.21$\pm$0.22 & 0.94$\pm$0.05  & 13.10$\pm$0.11  &  4.51$\pm$0.34  & 0.10$\pm$0.11  &BR/BL\\
18 09 00.0 & -20 03 34& G10.343-0.142 & M &3.82$\pm$0.27  & 12.20$\pm$0.04  &  2.14$\pm$0.10  & 2.15$\pm$0.27 & 1.61$\pm$0.49  & 12.00$\pm$0.08  &  2.21$\pm$0.26  & 1.03$\pm$0.80  &BR\\
18 09 01.8 & -20 05 10&G10.32-0.15 & M &3.70$\pm$0.24  & 12.40$\pm$0.03  &  2.09$\pm$0.08  & 1.95$\pm$0.24 & 2.44$\pm$0.59  & 12.10$\pm$0.06  &  1.81$\pm$0.16  & 2.08$\pm$0.80  &BR\\
18 09 03.5 & -20 02 54& G10.359-0.149A & MM & 1.20$\pm$0.01  & 11.50$\pm$0.06  &  2.95$\pm$0.12  & 0.10$\pm$0.04 & 0.56$\pm$0.01  & 11.10$\pm$0.13  &  2.98$\pm$0.12  & 0.11$\pm$0.05 &R \\
18 09 03.5 & -20 02 54& G10.359-0.149B & MM & 0.80$\pm$0.01  & 44.30$\pm$0.11  &  5.29$\pm$0.21  & 0.10$\pm$0.01 & 0.32$\pm$0.01  & 43.90$\pm$0.33  &  4.22$\pm$0.57  & 0.69$\pm$0.05 &R \\
18 10 15.6 & -19 54 45& G10.63-0.33B & MM & 3.02$\pm$0.23  & -4.72$\pm$0.05  &  2.34$\pm$0.09  & 2.40$\pm$0.32 & 2.41$\pm$0.58  & -4.81$\pm$0.11  &  1.92$\pm$0.19  & 4.43$\pm$1.30 &BR \\
18 10 18.4 & -19 54 29& G10.62-0.33 & M &4.54$\pm$0.21  & -4.41$\pm$0.03  &  2.47$\pm$0.07  & 2.48$\pm$0.21 & 1.85$\pm$0.40  & -4.65$\pm$0.08  &  2.76$\pm$0.21  & 1.42$\pm$0.62  &R\\
18 10 19.0 & -20 45 25 &G9.88-0.75 & R &6.69$\pm$0.25  & 28.30$\pm$0.02  &  2.00$\pm$0.04  & 2.46$\pm$0.16 & 2.16$\pm$0.41  & 28.10$\pm$0.05  &  2.38$\pm$0.16  & 0.83$\pm$0.47  &R\\
18 10 28.8 & -19 55 48& G10.62-0.38 & MR &1.51$\pm$0.04  & -2.79$\pm$0.06  &  4.04$\pm$0.13  & 0.10$\pm$0.06 & 0.98$\pm$0.04  & -4.65$\pm$0.13  &  5.46$\pm$0.23  & 0.10$\pm$0.14 &R/BL \\
18 10 41.1 & -19 57 41& G10.620-0.441 & MM & 3.32$\pm$0.24  & -1.30$\pm$0.03  &  1.68$\pm$0.07  & 1.73$\pm$0.26 & 0.72$\pm$0.08  & -1.67$\pm$0.10  &  2.30$\pm$0.27  & 0.10$\pm$0.74 &R  \\
18 11 23.9 & -19 32 20&  G11.075-0.384 & MM & 3.45$\pm$0.28  & -0.18$\pm$0.04  &  1.88$\pm$0.10  & 2.46$\pm$0.34 & $<$0.42  & - & - & - & BR\\
18 11 31.8 & -19 30 44& G11.11-0.34 & R &3.24$\pm$0.19  & 0.02$\pm$0.05  &  3.11$\pm$0.09  & 2.39$\pm$0.25 & 1.28$\pm$0.39  & 0.09$\pm$0.13  &  2.68$\pm$0.28  & 1.98$\pm$1.01 &R  \\
18 11 35.8 & -19 30 44& G11.117-0.413 & MM & 3.64$\pm$0.25  & -0.94$\pm$0.03  &  1.75$\pm$0.07  & 2.37$\pm$0.30 & 0.76$\pm$0.07  & -1.36$\pm$0.08  &  1.93$\pm$0.22  & 0.10$\pm$0.36 &R \\
18 11 51.4 & -17 31 30& G12.88+0.48 & M &2.83$\pm$0.17  & 33.00$\pm$0.05  &  2.94$\pm$0.10  & 1.63$\pm$0.21 & 1.47$\pm$0.35  & 33.00$\pm$0.10  &  3.33$\pm$0.29  & 1.36$\pm$0.67&R  \\
18 11 53.6 & -17 30 02& G12.914+0.493 & MM & 1.09$\pm$0.18  & 33.10$\pm$0.07  &  1.98$\pm$0.17  & 0.36$\pm$0.43 & 0.50$\pm$0.10  & 33.10$\pm$0.14  &  1.99$\pm$0.33  & 0.10$\pm$3.15 &R  \\
18 12 11.1 & -18 41 30& G11.903-0.140 & MR &2.71$\pm$0.20  & 38.50$\pm$0.07  &  3.39$\pm$0.14  & 2.42$\pm$0.33 & 0.53$\pm$0.05  & 38.20$\pm$0.21  &  4.98$\pm$0.60  & 0.10$\pm$0.88 &R \\
18 12 17.3 & -18 40 02& G11.93-0.14 & M &1.53$\pm$0.18  & 43.20$\pm$0.10  &  3.45$\pm$0.21  & 1.58$\pm$0.42 & $<$0.39  & - & - & - &R \\
18 12 33.1 & -18 30 05& G12.112-0.125 & MM & 1.01$\pm$0.17  & 45.20$\pm$0.09  &  2.13$\pm$0.24  & 0.98$\pm$0.48 & $<$0.18  & - & - & - &R\\
18 12 39.3 & -18 24 13& G12.20-0.09$^\dagger$ & MR &1.21$\pm$0.11  & 25.20$\pm$0.16  &  7.11$\pm$0.32  & 0.88$\pm$0.22 & 1.28$\pm$0.28  & 24.70$\pm$0.20  &  5.59$\pm$0.44  & 2.30$\pm$0.76 &R/BL \\
18 12 41.6 & -18 24 47& G11.942-0.256 & MM & 2.71$\pm$0.17  & 28.20$\pm$0.05  &  3.27$\pm$0.12  & 1.98$\pm$0.24 & 1.29$\pm$0.39  & 27.90$\pm$0.12  &  2.93$\pm$0.34  & 1.63$\pm$0.92 &R \\
18 12 43.3 & -18 25 09& G12.18-0.12A & M &2.51$\pm$0.16  & 28.10$\pm$0.05  &  3.42$\pm$0.11  & 1.66$\pm$0.23 & 0.80$\pm$0.05  & 27.80$\pm$0.12  &  3.89$\pm$0.29  & 0.10$\pm$0.30 &BR \\
18 12 44.4 & -18 24 21& G12.216-0.119 & MM & 1.77$\pm$0.16  & 28.50$\pm$0.06  &  2.85$\pm$0.16  & 1.13$\pm$0.28 & 0.67$\pm$0.06  & 28.30$\pm$0.12  &  3.55$\pm$0.39  & 0.10$\pm$0.45 &R \\
18 12 54.7 & -18 11 04& G12.43-0.05 & R &1.43$\pm$0.22  & 21.20$\pm$0.06  &  1.66$\pm$0.14  & 1.38$\pm$0.52 & 0.28$\pm$0.06 &-&-&- &BR \\ % & 20.70$\pm$0.32  &  3.10$\pm$0.94  & 0.10$\pm$1.00 &BR \\
18 13 54.1 & -18 01 41& G12.68-0.18 & M &6.10$\pm$0.22  & 55.80$\pm$0.03  &  2.88$\pm$0.06  & 3.11$\pm$0.18 & 1.87$\pm$0.35  & 55.60$\pm$0.06  &  3.19$\pm$0.23  & 0.89$\pm$0.46 &R \\
18 13 58.1 & -18 54 14& G11.94-0.62B & MM & 11.50$\pm$0.29  & 36.10$\pm$0.01  &  1.78$\pm$0.03  & 2.87$\pm$0.11 & 2.41$\pm$0.05  & 36.00$\pm$0.03  &  2.51$\pm$0.03  & 0.10$\pm$0.29 &R \\
18 14 00.9 & -18 53 18& G11.93-0.61 & MR &5.11$\pm$0.17  & 37.90$\pm$0.03  &  3.43$\pm$0.06  & 2.13$\pm$0.13 & 1.35$\pm$0.05  & 38.00$\pm$0.07  &  4.16$\pm$0.17  & 0.10$\pm$0.09 &R \\
18 14 07.0 & -18 00 37& G12.722-0.218 & MM & 1.15$\pm$0.12  & 34.40$\pm$0.10  &  4.07$\pm$0.23  & 0.88$\pm$0.31 & $<$0.31  & - & - & - &BR\\
18 14 24.9 & -17 53 44& G12.855-0.226 & MM & 9.47$\pm$0.32  & 36.30$\pm$0.01  &  1.57$\pm$0.03  & 3.08$\pm$0.17 & 3.31$\pm$0.60  & 36.10$\pm$0 .05  &  1.66$\pm$0.12  & 2.26$\pm$0.62&BR  \\
18 14 28.3& -17 52 00& G12.885-0.222 & MM & 6.46$\pm$0.27  & 36.30$\pm$0.02  &  2.08$\pm$0.05  & 2.39$\pm$0.17 & 1.37$\pm$0.42  & 36.20$\pm$0.09  &  2.24$\pm$0.24  & 1.10$\pm$0.87 &R \\
18 14 30.0 & -17 51 52& G12.892-0.226 & MM & 7.75$\pm$0.25  & 36.40$\pm$0.02  &  2.12$\pm$0.04  & 2.39$\pm$0.14 & 1.38$\pm$0.07  & 36.20$\pm$0.06  &  2.50$\pm$0.13  & 0.10$\pm$0.40  &R\\
18 14 33.9 & -17 51 44& G12.90-0.25B & MM & 11.60$\pm$0.31  & 37.10$\pm$0.01  &  2.04$\pm$0.03  & 2.69$\pm$0.12 & 2.24$\pm$0.07  & 36.90$\pm$0.04  &  2.75$\pm$0.10  & 0.10$\pm$0.12  &R\\
18 14 36.1 & -17 54 56& G12.859-0.272 & MM & 3.23$\pm$0.19  & 36.70$\pm$0.05  &  3.19$\pm$0.12  & 1.53$\pm$0.20 & 1.89$\pm$0.54  & 36.50$\pm$0.10  &  2.14$\pm$0.24  & 2.08$\pm$0.95 &BR \\
18 14 35.5 & -16 45 36& G13.87+0.28 & M &0.63$\pm$0.07  & 48.80$\pm$0.11  &  2.33$\pm$0.33  & 0.10$\pm$0.35 & 0.56$\pm$0.12  & 48.70$\pm$0.14  &  2.18$\pm$0.37  & 0.10$\pm$19.10 &R \\
18 14 38.9 & -17 51 52& G12.90-0.26 & M &7.88$\pm$0.23  & 36.80$\pm$0.02  &  3.84$\pm$0.06  & 1.82$\pm$0.10 & 2.82$\pm$0.43  & 36.80$\pm$0.05  &  3.76$\pm$0.21  & 0.49$\pm$0.34  &R\\
18 14 41.7 & -17 54 24& G12.878-0.226 & MM & 4.00$\pm$0.19  & 34.60$\pm$0.05  &  3.82$\pm$0.09  & 1.92$\pm$0.18 & $<$0.37  & - & - & - &R\\
18 14 42.9& -17 53 12& G12.897-0.281 & MM & 3.62$\pm$0.18  & 35.50$\pm$0.06  &  4.54$\pm$0.10  & 1.98$\pm$0.18 & 1.04$\pm$0.34  & 35.30$\pm$0.17  &  4.42$\pm$0.52  & 1.04$\pm$0.86 &R/BL \\
18 14 44.5 & -17 52 16& G12.914-0.280 & MM & 2.46$\pm$0.18  & 35.40$\pm$0.10  &  4.85$\pm$0.18  & 2.17$\pm$0.28 & $<$0.29  & - & - & - &R/BL \\
18 14 45.7 & -17 50 48& G12.938-0.272 & MM & 4.17$\pm$0.27  & 34.80$\pm$0.04  &  2.31$\pm$0.11  & 2.19$\pm$0.25 & $<$0.51  & - & - & - &R\\
18 16 22.1 & -19 41 19& G11.49-1.48 & M &2.08$\pm$0.29  & 10.60$\pm$0.06  &  1.54$\pm$0.13  & 1.70$\pm$0.52 & $<$0.48  & - & - & - &BR\\
18 17 02.2 & -16 14 28& G14.60+0.01 & MR &2.37$\pm$0.19  & 25.10$\pm$0.07  &  3.12$\pm$0.14  & 1.97$\pm$0.31 & 1.69$\pm$0.43  & 25.20$\pm$0.14  &  3.35$\pm$0.33  & 2.36$\pm$0.90 &R \\
18 19 12.0 & -20 47 23& G10.84-2.59 & R &1.05$\pm$0.22  & 12.50$\pm$0.07  &  1.74$\pm$0.20  & 0.39$\pm$0.52 & 0.32$\pm$0.06 &-&-&-& R\\% & 12.40$\pm$0.25  &  2.53$\pm$0.46  & 0.10$\pm$0.40 &R \\
18 21 14.6 & -14 32 52& G16.580-0.079 & MM & 2.35$\pm$0.33  & 41.10$\pm$0.05  &  1.35$\pm$0.12  & 2.46$\pm$0.59 & $<$0.48  & - & - & - &R\\
18 21 09.1 & -14 31 40& G16.58-0.05 & M &2.38$\pm$0.20  & 60.00$\pm$0.04  &  2.31$\pm$0.10  & 1.39$\pm$0.27 & 0.73$\pm$0.06  & 59.80$\pm$0.10  &  2.60$\pm$0.22  & 0.10$\pm$0.28&R  \\
18 25 41.7 & -13 10 16&G18.30-0.39 & R &2.34$\pm$0.20  & 32.90$\pm$0.04  &  1.89$\pm$0.09  & 0.82$\pm$0.24 & 1.49$\pm$0.46  & 32.70$\pm$0.07  &  1.85$\pm$0.22  & 0.96$\pm$0.79 &R \\
18 27 16.3 & -11 53 51& G19.61-0.1 & M &0.67$\pm$0.12  & 59.50$\pm$0.19  &  5.29$\pm$0.45  & 0.24$\pm$0.36 & 0.41$\pm$0.04  & 59.00$\pm$0.30  &  6.33$\pm$0.62  & 0.10$\pm$0.22  &BR/BL \\
18 29 24.2 & -15 15 34& G16.871-2.154 & MM & 7.91$\pm$0.18  & 19.20$\pm$0.02  &  2.99$\pm$0.04  & 2.03$\pm$0.09 & 2.57$\pm$0.32  & 19.10$\pm$0.04  &  3.50$\pm$0.16  & 0.55$\pm$0.28 &R \\
18 29 24.2 & -15 16 06& G16.86-2.15 & M &3.82$\pm$0.19  & 18.40$\pm$0.04  &  2.97$\pm$0.08  & 1.86$\pm$0.19 & 1.08$\pm$0.06  & 18.00$\pm$0.09  &  3.70$\pm$0.22  & 0.10$\pm$0.99 &R \\
18 31 43.0 & -09 22 28& G22.36+0.07B & M &2.87$\pm$0.27  & 84.70$\pm$0.04  &  1.72$\pm$0.10  & 2.09$\pm$0.37 & $<$0.48  & - & - & -&R \\
18 33 53.1 & -08 07 23& G23.71+0.17 & R &1.23$\pm$0.14  & 113.00$\pm$0.17  &  5.15$\pm$0.33  & 1.22$\pm$0.35 & 0.53$\pm$0.06  & 113.00$\pm$0.17  &  3.33$\pm$0.40  & 0.10$\pm$0.18 &BR/BL   \\
18 33 53.6 & -08 08 51& G23.689+0.159 & MM & 0.56$\pm$0.07  & 113.00$\pm$0.17  &  3.94$\pm$0.43  & 0.10$\pm$6.41 & 0.30$\pm$0.05  & 112.00$\pm$0.53  &  6.41$\pm$1.40  & 0.10$\pm$0.42 &BR \\
18 34 39.2 & -08 31 41& G23.43-0.18 & M &2.56$\pm$0.17  & 101.00$\pm$0.08  &  4.63$\pm$0.16  & 1.34$\pm$0.21 & 0.78$\pm$0.05  & 101.00$\pm$0.22  &  7.68$\pm$0.62  & 0.10$\pm$0.21 &R/BL  \\
18 34 36.2 & -08 42 39&G23.268-0.257A$^\ast$ & MM & 1.20$\pm$0.01  & 61.30$\pm$0.07  &  2.83$\pm$0.14  & 0.10$\pm$0.03 & 0.48$\pm$0.04  & 60.80$\pm$0.19  &  2.94$\pm$0.43  & 0.10$\pm$0.25 &BR \\ 
18 34 45.7 & -08 34 21&G23.409-0.228 & MM & 2.34$\pm$0.19  & 104.00$\pm$0.10  &  4.12$\pm$0.20  & 1.57$\pm$0.29 & 0.74$\pm$0.08  & 104.00$\pm$0.13  &  2.50$\pm$0.36  & 0.10$\pm$0.72 &BR \\
18 36 06.7 & -07 13 47& G23.754+0.095 & MM & 2.09$\pm$0.31  & 110.00$\pm$0.08  &  2.27$\pm$0.23  & 1.32$\pm$0.47 & $<$0.66  & - & - & - &BR \\
18 46 01.3 & -02 45 25& G29.861-0.053$^\dagger$ & MM & 0.96$\pm$0.20  & 99.90$\pm$0.11  &  2.41$\pm$0.35  & 0.85$\pm$0.62 & 2.80$\pm$0.92  & 100.00$\pm$0.17  &  1.83$\pm$0.26  & 9.39$\pm$3.70 &BR \\
18 46 03.9 & -02 39 25& G29.96-0.02B & MR &1.42$\pm$0.17  & 98.00$\pm$0.14  &  4.57$\pm$0.31  & 1.07$\pm$0.36 & 0.80$\pm$0.09  & 97.40$\pm$0.14  &  2.91$\pm$0.43  & 0.10$\pm$0.67 &BR/BL  \\
18 46 05.0 & -02 42 29& G29.912-0.045 & MM & 2.54$\pm$0.20  & 101.00$\pm$0.10  &  4.49$\pm$0.18  & 1.81$\pm$0.28 & 0.80$\pm$0.05  & 100.00$\pm$0.19  &  6.45$\pm$0.48  & 0.10$\pm$0.15 &BR/BL \\
\hline
\end{tabular}
%\end{table*}
%\newpage 
%\addtocounter{table}{-1} 

%\begin {table*}
%\caption{Cont.}
% \begin{tabular}{@{}l|@{}l|@{}l|@{}c|@{}c|@{}c|@{}c|@{}c|@{}c|@{}c|@{}c|@{}c|@{}c|}  \hline 
%Right & Declin- &Source & Class$^\alpha$ & Flux Den.$^{1,1}$ & V$_{\rm LSR}^{1,1}$ & $\Delta$V$^{1,1}$ & $\tau_{\rm main}^{1,1}$ & Flux Den.$^{2,2}$ & V$_{\rm LSR}^{2,2}$ & $\Delta$V$^{2,2}$ & $\tau_{\rm main}^{2,2}$ & Base- \\
%Ascension & ation&       & & (Jy/beam)     & (km/s)      & (km/s)    &    &  (Jy/beam)     & (km/s) & (km/s)       & &Line$^\beta$   \\ \hline
%
%\hline
%\end{tabular}

\begin{flushleft}
$^\alpha$ Denotes the source class, with MM indicative of MM-only sources, M for methanol maser associations, R for radio continuum associations and MR for sources with both a methanol maser and radio continuum source. \\
$^\beta$ Indicates the reliability of the baseline, with R=Reliable; BR=Baseline ripple and BL= blended. See section~\ref{sec:fitting}.\\
$^\dagger$ Denotes sources where the \nhtwo\, Flux Density is greater than the \nhone.\\
$^\ast$ Indicates that the second component of this spectrum (Fig~\ref{fig:spectra}) could not be fit.
\end{flushleft}
\end{table*}

Column densities and rotational temperatures were derived in the standard way \citep[e.g.][]{ungerechts86}. The kinetic temperature estimates were calculated from the rotational temperature following the procedure outlined in \citet{tafalla04}. The derived molecular gas properties for each source are presented in Table~\ref{tab:nh3_derivs}, in right ascension order. The rotational temperature (\trot) and kinetic temperature (\tkin) are given in columns 2 and 5, with the lower and upper limits to each temperature given in columns 3 \& 4 and 6 \& 7, respectively. Temperature uncertainties are discussed further below. The column density of the \nhone (N$_{\nhone}$) and the total column density of the gas (N$_{\nh}^{TOT}$) are presented in columns 8 and 9, respectively. 

Given the varying robustness of the \nhone\, and (2,2) detections, sources were divided into 4 groups based on the reliability of the spectra (and hence derived physical properties). Group 4 sources were defined as those with no \nhone\, detection. No physical properties were derived for these sources, which can be found in Table~\ref{tab:non_detections}. Sources with an \nhone\, detection, but with either i) no \nhtwo\, detection or ii) the \nhone\, \& (2,2) \vlsr/\delv\, differed by $>$3 km/s (making it unlikely the emission is coming from the same gas) were defined as group 3. For these sources, there is no \nhtwo\, emission at the position of the \nhone\, detection, so the RMS of the \nhtwo\, spectra was used to derive an upper temperature limit and column density estimate. These numbers are highly uncertain. Group 2 sources are defined as those with both \nhone\, and (2,2) detections which have consistent kinematics but the signal to noise of at least one spectra lies in the range 3 $<$ $\sigma$ $<$ 10. Finally, sources with both \nhone\, and (2,2) detections $>$10$\sigma$ are defined as group 1. These group allocations are given in column 10 of Table~\ref{tab:nh3_derivs}.

Uncertainties in the temperature were estimated for group 1 and 2 sources using the uncertainties in the measured parameters (Flux density, $\vlsr$, $\delv$ \& $\tau$) from the fits to the spectra. For each core, a maximum and minimum rotational temperature were derived assuming each of the parameters was at 1$\sigma$ above or below the actual measured value (e.g. Flux$^{11}$ $+ \Delta$Flux$^{11}$ or Flux$^{11}$ $- \Delta$Flux$^{11}$). The lower and upper rotational temperatures are shown in columns 3 and 4 of Table~\ref{tab:nh3_derivs}. These extrema were then used to derive corresponding lower and upper limits to the kinetic temperatures, again following \citet{tafalla04}. The lower and upper kinetic temperatures are shown in columns 6 and 7 of Table~\ref{tab:nh3_derivs}.  As a general rule, the cores at low temperatures are well constrained, while those above $\sim$20\,K are poorly constrained. This is reflected in the larger error bars for poorly constrained temperatures, and is not unexpected given the insensitivity of \nhone\, and (2,2) as a temperature probe of warmer gas \citep[e.g.][]{danby88}.

For some sources (e.g. G305.776--0.251) the rotational temperature was so high that the analytic for determined by \citet{tafalla04} was no longer reliable. These sources have very small ($<$8K) lower limits, and no upper limit could be derived.

\begin {table*}
\caption{Parameters derived from the \nhone\, and \nhtwo\, spectra. Sources are in right ascension order, as per Table~\ref{tab:nh3_11_22_fits}.}
\label{tab:nh3_derivs} 
\begin{tabular}{|l|c|c|c|c|c|c|c|c|c|} \hline 
Source & T$_{\rm rot}$$^a$ &T$^{\rm lower}_{\rm rot}$ &  T$^{\rm upper}_{\rm rot}$ & T$_{\rm kin}$$^a$ & T$^{\rm lower}_{\rm kin}$ &  T$^{\rm upper}_{\rm kin}$ & N$_{NH3(1,1)}$ & N$_{NH3}^{TOT}$ & Rel.\\ 
       & (K)              & (K)              & (K)              & (K)              &   (K)              & (K)              & (10$^{13}$cm$^{-2}$)    & (10$^{14}$cm$^{-2}$)  & Group$^\beta$    \\ \hline 
 G269.15-1.13 &   28 &   18 &   63 &  42 &   22 &  -   & 5.9 &  0.7 &  1 \\
 G291.256-0.769 &   16 &    9 &   39 &  19 &    9 &   82 & 8.6 &  1.0 &  1 \\
 G291.256-0.743 &   22 &   19 &   28 &  29 &   22 &   41 & 8.3 &  0.9 &  1 \\
 G291.309-0.681 &   23 &   15 &   53 &  31 &   17 &  - & 4.8 &  0.5 &  2 \\
 G291.576-0.468 & $<$  22 & - & - &$<$  22 & - & -       & 4.5 &  0.5 &  3 \\
 G291.58-0.53 & $<$  23 & - & - &$<$  23 & - & -         & 4.9 &  0.6 &  3 \\
 G294.97-1.7 & $<$  16 & - & - &$<$  16 & - & -          & 4.3 &  0.5 &  3 \\
G304.890+0.636 & $<$   9 & - & - &$<$   9 & - & -        & 2.5 &  0.7 &  3 \\  
G305.145+0.208 & $<$  14 & - & - &$<$  14 & - & -        & 4.3 &  0.6 &  3 \\
 G305.137+0.069 & $<$  12 & - & - &$<$  12 & - & -       & 9.7 &  1.7 &  3 \\
 G305.192-0.006 &   20 &   12 &   45 &  25 &   13 &  119 &11.7 &  1.3 &  1 \\
 G305.21+0.21 &   37 &   26 &   66 &  71 &   37 & -   &12.2 &  1.6 &  2 \\
 G305.197+0.007 &   14 &   10 &   22 &  16 &   11 &   28 &15.6 &  2.1 &  2 \\
 G305.226+0.275 &   23 &   18 &   32 &  31 &   22 &   51 &18.7 &  2.1 &  2 \\
 G305.228+0.286 &   35 &   23 &   66 &  62 &   30 & - &16.6 &  2.0 &  2 \\
 G305.238+0.261 &   18 &   12 &   31 &  21 &   13 &   48 &10.4 &  1.2 &  2 \\
 G305.248+0.245 & $<$  13 & - & - &$<$  13 & - & -       & 8.1 &  1.2 &  3 \\
 G305.233-0.023 &   16 &   12 &   24 &  19 &   13 &   32 &16.8 &  2.1 &  2 \\
 G305.269-0.010 &   15 &   13 &   17 &  17 &   15 &   20 &19.1 &  2.5 &  1 \\
 G305.355+0.194 &   21 &   13 &   44 &  27 &   15 &  106 &13.2 &  1.5 &  2 \\
 G305.37+0.21 & $<$  16 & - & - &$<$  16 & - & -         &13.4 &  1.7 &  3 \\
 G305.362+0.185 &   22 &   16 &   35 &  28 &   18 &   61 &17.8 &  2.0 &  2 \\
 G305.361+0.151 & $<$  26 & - & - &$<$  26 & - & -       & 2.5 &  0.3 &  3 \\
 G305.538+0.340 & $<$  14 & - & - &$<$  14 & - & -       &10.3 &  1.4 &  3 \\
 G305.55+0.01 &   23 &   13 &   79 &  30 &   14 & -      & 5.4 &  0.6 &  2 \\
 G305.552+0.012 & $<$  26 & - & - &$<$  26 & - & -       & 2.0 &  0.2 &  3 \\
 G305.561+0.012 &   29 &   18 &   80 &  44 &   21 & -    & 3.2 &  0.3 &  2 \\
G305.776-0.251 &   17 &    2 &  - &  19 &    2 & -     & 4.0 &  0.5 &  2 \\ 
G305.81-0.25 &   22 &   16 &   37 &  28 &   18 &   69    & 9.4 &  1.0 &  2 \\
 G305.833-0.196 &   12 &    7 &   26 &  13 &    7 &   37 & 6.1 &  1.0 &  2 \\
 G306.33-0.3 & $<$   8 & - & - &$<$   8 & - & -          & 6.5 &  2.0 &  3 \\
 G306.343-0.302 & $<$  19 & - & - &$<$  19 & - & -       & 8.4 &  1.0 &  3 \\
 G309.917+0.494 & $<$  14 & - & - &$<$  14 & - & -       & 8.6 &  1.2 &  3 \\
 G309.92+0.4 & $<$  24 & - & - &$<$  24 & - & -          & 3.7 &  0.4 &  3 \\
 G318.92-0.68 &   24 &   16 &   43 &  33 &   19 &  104   &13.0 &  1.4 &  2 \\
 G323.74-0.3 &   30 &   16 &  141 &  46 &   18 & -       & 8.0 &  0.9 &  2 \\
 G332.646-0.647A &   17 &   14 &   20 &  19 &   16 &  24 & 6.3 &  0.7 &  2 \\
 G332.646-0.647B &   20 &   17 &   25 &  25 &   19 &  35 & 3.1 &  0.3 &  2 \\
 G332.695-0.609 & $<$  10 & - & - &$<$  10 & - & -       &15.6 &  3.1 &  3 \\
 G332.725-0.62 & $<$  10 & - & - &$<$  10 & - & -        & 9.2 &  2.1 &  3 \\
 G332.627-0.511 &   17 &   12 &   28 &  19 &   13 &   41 & 6.4 &  0.8 &  2 \\
 G332.827-0.552 &   18 &   12 &   32 &  22 &   13 &   52 &16.0 &  1.9 &  1 \\
 G0.331-0.164 & $<$   7 & - & - &$<$   7 & - & -         &12.3 &  4.9 &  3 \\
 G0.310-0.170 & $<$  10 & - & - &$<$  10 & - & -         &11.7 &  2.6 &  3 \\
 G0.32-0.20 &   23 &   15 &   38 &  29 &   18 &   74     &11.0 &  1.3 &  2 \\
 G1.105-0.098 &   32 &   27 &   40 &  53 &   39 &   83   & 5.8 &  0.7 &  1 \\
 G1.13-0.11 &   30 &   28 &   34 &  47 &   40 &   58     & 6.3 &  0.7 &  1 \\
 G0.549-0.868 &   16 &   13 &   20 &  18 &   14 &   25   & 6.5 &  0.9 &  1 \\
 G0.627-0.848 & $<$  11 & - & - &$<$  11 & - & -         &13.4 &  2.4 &  3 \\
 G0.600-0.871 &   13 &    5 &   53 &  14 &    5 &  221   & 9.7 &  1.4 &  2 \\
 G2.54+0.20 &   12 &    9 &   19 &  13 &    9 &   23     &10.6 &  1.7 &  1 \\
 G5.504-0.246 &   12 &    9 &   16 &  13 &    9 &   19   & 9.4 &  1.6 &  2 \\
 G5.90-0.42 &   16 &   12 &   23 &  18 &   13 &   30     &18.9 &  2.4 &  2 \\
 G5.90-0.44 &   23 &   16 &   40 &  30 &   18 &   84     &10.2 &  1.1 &  2 \\
 G8.111+0.257 &   29 &   18 &   76 &  44 &   21 & -      & 1.8 &  0.2 &  2 \\
G8.127+0.255 &   17 &   13 &   27 &  20 &   14 &   38    & 7.2 &  0.9 &  2 \\
 G8.138+0.246 &   17 &   13 &   26 &  20 &   14 &   36   & 8.2 &  1.0 &  2 \\
G8.13+0.22 &   15 &   12 &   18 &  16 &   13 &   21      &11.9 &  1.6 &  1 \\ 
 G5.948-1.125 & $<$   9 & - & - &$<$   9 & - & -         & 2.7 &  0.7 &  3 \\
 G5.962-1.128 &   13 &    4 &  - &  14 &    4 & -      & 5.2 &  0.8 &  2 \\
 G5.975-1.146 &   21 &   15 &   35 &  26 &   17 &   61   & 6.1 &  0.7 &  2 \\
 G5.971-1.158 & $<$  12 & - & - &$<$  12 & - & -         & 5.3 &  0.9 &  3 \\
 G9.63+0.19 &   27 &   19 &   44 &  38 &   22 &  110     &19.2 &  2.2 &  2 \\
 G8.68-0.36 &   12 &   11 &   14 &  13 &   11 &   16     &38.9 &  6.2 &  2 \\
\end{tabular}
\end{table*}
%\newpage 
\addtocounter{table}{-1} 
\begin {table*}
\caption{Cont.}
\begin{tabular}{|l|c|c|c|c|c|c|c|c|c|} \hline 
Source & T$_{\rm rot}$$^a$ & T$^{\rm lower}_{\rm rot}$  & T$^{\rm upper}_{\rm rot}$ & T$_{\rm kin}$$^a$ & T$^{\rm lower}_{\rm kin}$ &  T$^{\rm upper}_{\rm kin}$  & N$_{NH3(1,1)}$ & N$_{NH3}^{TOT}$ & Rel.\\
       & (K)              & (K)              & (K)              & (K)              &   (K)              & (K)              & (10$^{13}$cm$^{-2}$)    & (10$^{14}$cm$^{-2}$)  & Group$^\beta$    \\ \hline
 G8.686-0.366 &   16 &   13 &   20 &  19 &   15 &   24  &35.6 &  4.5 &  2 \\
 G8.644-0.395 &   11 &    8 &   15 &  11 &    8 &   17  & 9.9 &  1.9 &  2 \\
 G8.713-0.364 &    9 &    8 &   10 &   9 &    8 &   11  &23.3 &  6.2 &  1 \\
 G8.735-0.362 &   12 &    9 &   17 &  13 &   10 &   20  &17.6 &  2.8 &  2 \\
 G8.724-0.401 &   10 &    7 &   13 &  10 &    7 &   14  &20.1 &  4.6 &  1 \\
 G8.709-0.412 &   11 &    9 &   14 &  12 &    9 &   16  &15.9 &  2.9 &  1 \\
 G8.718-0.410 &   13 &   10 &   16 &  14 &   11 &   18  &19.8 &  3.0 &  2 \\
 G10.287-0.110 &   22 &   15 &   42 &  29 &   16 &   93 &18.9 &  2.1 &  2 \\
 G10.284-0.126 &   25 &   17 &   40 &  33 &   20 &   86 &17.4 &  2.0 &  2 \\
 G10.288-0.127 &   19 &   16 &   25 &  23 &   18 &   34 &10.7 &  1.3 &  1 \\
 G10.29-0.14 &   17 &   14 &   21 &  20 &   16 &   26   &15.5 &  1.9 &  1 \\
 G10.343-0.142 &   16 &   11 &   27 &  19 &   12 &   38 &14.3 &  1.8 &  2 \\
 G10.32-0.15 &   24 &   16 &   45 &  33 &   18 &  120   &13.0 &  1.4 &  2 \\
 G10.359-0.149A &   26 &   23 &   29 &  37 &   31 &  44 & 3.3 &  0.3 &  1 \\
 G10.359-0.149B &   25 &   22 &   29 &  35 &   28 &  45 & 4.2 &  0.5 &  1 \\
 G10.63-0.33B &   38 &   20 &  157 &  75 &   25 & -     &15.1 &  1.9 &  2 \\
 G10.62-0.33 &   17 &   13 &   24 &  20 &   14 &   31   &19.2 &  2.4 &  2 \\
 G10.62-0.38 &   42 &   33 &   61 &  98 &   54 &  -   & 5.5 &  0.7 &  1 \\
G9.88-0.75 &   14 &   12 &   18 &  16 &   12 &   21     &18.8 &  2.6 &  2 \\
 G10.620-0.441 &   13 &   10 &   18 &  14 &   10 &   21 & 9.3 &  1.4 &  2 \\
 G11.075-0.384 & $<$   9 & - & - &$<$   9 & - & -       &12.9 &  3.2 &  3 \\
 G11.11-0.34 &   17 &   12 &   27 &  20 &   12 &   38   &20.6 &  2.5 &  2 \\
 G11.117-0.413 &   11 &    9 &   13 &  12 &   10 &  14  &12.1 &  2.2 &  1 \\
 G12.88+0.48 &   23 &   15 &   40 &  30 &   17 &   84   &14.7 &  1.7 &  2 \\
G12.914+0.493 &   22 &    8 &  - &  29 &    8 & -     & 2.9 &  0.3 &  2 \\ 
G11.903-0.140 &   12 &    9 &   16 &  12 &    9 &   18  &21.2 &  3.6 &  1 \\
 G11.93-0.14 & $<$  12 & - & - &$<$  12 & - & -         &13.4 &  2.1 &  3 \\
 G12.112-0.125 & $<$  12 & - & - &$<$  12 & - & -       & 5.2 &  0.8 &  3 \\
 G11.942-0.256 &   19 &   12 &   33 &  23 &   13 &   55 &18.0 &  2.1 &  2 \\
 G12.18-0.12A &   14 &   12 &   17 &  15 &   13 &   19  &16.4 &  2.3 &  1 \\
 G12.216-0.119 &   17 &   13 &   25 &  20 &   14 &   35 & 9.4 &  1.1 &  1 \\
 G12.43-0.05 &   15 &    9 &   34 &  16 &    9 &   59   & 5.7 &  0.7 &  2 \\
 G12.68-0.18 &   13 &   11 &   16 &  14 &   11 &   18   &29.0 &  4.4 &  2 \\
 G11.94-0.62B &   11 &   10 &   12 &  12 &   11 &   13  &24.7 &  4.5 &  1 \\
 G11.93-0.61 &   12 &   12 &   13 &  13 &   12 &   15   &26.1 &  4.1 &  1 \\
 G12.722-0.218 & $<$  15 & - & - &$<$  15 & - & -       & 9.8 &  1.3 &  3 \\
 G12.855-0.226 &   16 &   13 &   21 &  19 &   14 &   27 &19.8 &  2.5 &  2 \\
 G12.885-0.222 &   13 &   10 &   18 &  14 &   10 &   21 &19.0 &  2.9 &  2 \\
 G12.892-0.226 &   11 &    9 &   12 &  11 &   10 &   13 &21.4 &  4.1 &  1 \\
 G12.90-0.25B &   11 &   10 &   12 &  12 &   11 &   12  &27.8 &  5.2 &  1 \\
 G12.859-0.272 &   22 &   14 &   42 &  28 &   16 &   97 &16.4 &  1.8 &  2 \\
 G13.87+0.28 &   40 &    2 &  - &  86 &    2 & -      & 1.6 &  0.2 &  2 \\
 G12.90-0.26 &   14 &   12 &   17 &  16 &   13 &   20   &35.5 &  4.8 &  2 \\
 G12.878-0.226 & $<$   9 & - & - &$<$   9 & - & -       &24.4 &  6.0 &  3 \\
 G12.897-0.281 &   14 &   10 &   21 &  16 &   11 &   27 &28.2 &  3.9 &  2 \\
 G12.914-0.280 & $<$   9 & - & - &$<$   9 & - & -       &27.3 &  6.4 &  3 \\
 G12.938-0.272 & $<$   9 & - & - &$<$   9 & - & -       &16.1 &  3.7 &  3 \\
 G11.49-1.48 & $<$  12 & - & - &$<$  12 & - & -         & 7.0 &  1.1 &  3 \\
 G14.60+0.01 &   31 &   18 &   96 &  50 &   21 & -      &16.3 &  1.9 &  2 \\
 G10.84-2.59 &   22 &   12 &   86 &  27 &   14 & -      & 2.6 &  0.3 &  2 \\
 G16.580-0.079 & $<$  10 & - & - &$<$  10 & - & -       & 8.2 &  1.6 &  3 \\
 G16.58-0.05 &   14 &   12 &   18 &  16 &   13 &   22   & 9.8 &  1.3 &  1 \\ 
G18.30-0.39 &   29 &   15 &  109 &  43 &   18 & -       & 6.1 &  0.7 &  2 \\
 G19.61-0.1 &   32 &   19 &  122 &  52 &   23 & -       & 4.9 &  0.6 &  2 \\
 G16.871-2.154 &   14 &   13 &   17 &  16 &   14 &   19 &28.8 &  3.9 &  2 \\
 G16.86-2.15 &   13 &   10 &   18 &  14 &   11 &   21   &18.3 &  2.7 &  1 \\
 G22.36+0.07B & $<$  10 & - & - &$<$  10 & - & -        &10.0 &  2.0 &  3 \\
 G23.71+0.17 &   14 &   11 &   18 &  15 &   12 &   22   &15.7 &  2.2 &  2 \\
 G23.43-0.18 & $<$   7 & - & - &$<$   7 & - & -         &19.8 &  9.0 &  3 \\
G23.268-0.257A &   24 &   19 &   33 &  32 &   22   & 55 & 3.1 &  0.3 &  3 \\ 
G23.409-0.228 &   12 &    9 &   16 &  12 &    9 &   18  &18.6 &  3.2 &  2 \\
 G23.754+0.095 & $<$  14 & - & - &$<$  14 & - & -       & 8.8 &  1.2 &  3 \\
 G29.96-0.02B &   16 &   11 &   26 &  18 &   12 &   37  &13.4 &  1.7 &  2 \\
 G29.912-0.045 &   14 &   12 &   18 &  16 &   13 &   21 &22.8 &  3.0 &  1 \\
\hline
\end{tabular}
\begin{flushleft}
$^a$ Note that \nhone\, and \nhtwo\, are only sensitive to temperatures below 30\,K (see sections~\ref{sec:tkin} and \ref{sec:dis:sed}).\\
$^\beta$ Refers to the robustness of the detection, as outlined in section~\ref{sec:fitting}.\\
\end{flushleft}
\end{table*}
%\addtocounter{table}{-1}
\begin{table}
\caption{Sources which were observed in \nhone\, and (2,2) but are non-detections (group 4 sources) or are too confused to fit.}
 \label{tab:non_detections} 
\scriptsize
\begin{tabular}{|lp{0.3cm}clp{0.3cm}c} \hline
Source Name & Class$^a$ & $^{b,c}$ &\vline~ Source Name & Class$^a$ & $^{b,c}$ \\
\hline
G269.45-1.47 &  MR & ND &\vline~ G305.519-0.040	& MM & ND \\  
G270.25+0.84 & M  & ND &\vline~ G305.520-0.020	& MM & ND \\	
G284.271-0.391 & MM & ND &\vline~ G305.549+0.002	& MM & M\\  %wrong velocity for this region? & & % wek detection, stable baseline\\
G284.295-0.362 & MM & ND &\vline~ G305.581+0.033 & MM& ND\\
G284.307-0.376 & MM & ND &\vline~ G305.605+0.010	& MM & M/ND\\ %& &% noisy spectrum, possible something at position 
G284.338-0.417 & MM & ND &\vline~ G306.319-0.343	& MM & ND \\
G284.35-0.42 & M & M/ND &\vline~ G318.913-0.162	& R & ND \\  %only maser in region
G284.345-0.404 & MM & ND &\vline~ G330.952-0.18	& MR & M \\  % && appear to be something, ut not strong enough.
G284.341-0.389 & MM & M/ND &\vline~ G332.640-0.586	& MM & D/M\\ % something slithgly stronger than noise at 3km/s && % too weaek
G284.328-0.365 & MM& ND &\vline~ G332.701-0.587	& MM & ND\\
G284.384-0.441 & MM & M/ND &\vline~ G332.777-0.584	& MM & M/ND \\ % something at position, but same strength as the noise at 3km/s. && % strong suggestion of soemthing, but lost in the spectrum and the noise\\
G284.344-0.366 & MM & ND &\vline~ G332.794-0.598	& MM & ND \\
G284.352-0.353 & MM & ND &\vline~ G0.204+0.051 & MM & SB \\
G287.37+0.65 & M & ND &\vline~ G0.49+0.19& M & SA$^\gamma$\\  
G290.40-2.91 & M & ND&\vline~ G0.266-0.034& MM & SB\\
G291.27-0.70 & MR & D  &\vline~ G0.21-0.00 & MR & SB\\ % too weak compared to noise
G291.288-0.706 & MM & ND &\vline~ G0.497+0.170 & MM & SA$^\gamma$\\
G291.302-0.693 & MM & M &\vline~ G0.240+0.008 &  MM &SB\\ %something weak at rest velocity
G290.37+1.66 & M & M/ND &\vline~ G0.527+0.181 & R & SA$^\gamma$\\	%something weak at rest velocity?
G291.587-0.499 & MM & ND &\vline~ G0.271+0.022 & MM & SB\\ 	
G291.572-0.450 & MM & M/ND &\vline~ G0.257+0.011 & MM & SB\\% && incredibly strong and blended
G291.608-0.532 & MM & ND &\vline~ G0.83+0.18& M & M/ND\\ % &&noisy
G291.597-0.496 & MM & ND &\vline~ G0.325-0.242& MM & ND\\
G291.630-0.545 & MM & ND &\vline~ G1.124-0.065 	& MM & SB	\\
G291.614-0.443 & MM & ND &\vline~ G1.134-0.073  &MM & SB	\\
G293.824-0.762 & MM & ND &\vline~ G1.14-0.12 & M & S$^\gamma$\\% &&GP second source, unstable baseline, ubut definite strong detection. Fit wont converge due to baseline \\ \\
G293.82-0.74& MR & M &\vline~ G0.55-0.85& MR & SB\\ % something weak at position && blended   	
G293.892-0.782	& MM& D/ND &\vline~ G5.48-0.24 & R	& ND \\
G293.95-0.8	& MR & ND &\vline~ G6.53-0.10 & R & M/ND	\\ % &&soemthing in spectrum but too weak\\
G293.942-0.876 & MM & ND &\vline~ G6.60-0.08 & M	& M/ND\\ %&& soething in spectum but weak
G293.989-0.936	& MM & M/ND &\vline~ G6.62-0.10 & MM & ND	\\ %something weak at point, but bad ripples
G294.52-1.6& M	& M/ND &\vline~ G5.97-1.17 & R	& ND \\% something weak maybe, but baseline unstable
G294.945-1.737	& MM & ND &\vline~ G10.10+0.72& R & ND\\
G294.989-1.720 & M & ND &\vline~ G9.966-0.020 & MM & M/ND\\ % &&stable baseline too.\\
G298.26+0.7& M	& D/ND &\vline~ G9.99-0.03& M & ND\\ % obvious detection at position but too weak
G299.02+0.1& M & ND&\vline~ G10.001-0.033 & R & ND\\
G299.024+0.130 & MM & ND &\vline~ G10.44-0.01 & M & D/ND\\ % &&too weak
G300.455-0.190	& MM & ND&\vline~ G9.924-0.749& MM & ND 		\\
G300.51-0.1 & M & ND &\vline~ G11.948-0.003 & MM& ND\\
G301.14-0.2 & MR & ND &\vline~ G12.02-0.03 & M  & D/ND\\% && too weak
G302.03-0.06 & MR & D/ND &\vline~ G11.902-0.100 & MM &M/ND\\ % % weak detection at position &&suggestions of somethin, but too weak	
G304.906+0.574	& MM & M/ND &\vline~ G11.861-0.183 & MM & ND\\ %  detection but spectrum is noisy
G304.919+0.542	& MM & ND &\vline~ G11.942-0.157 & MM& D/M	\\ % && too weak
G305.952+0.555	& MM & ND &\vline~ G12.200-0.003& MM  & D/M	\\ %&& too weak\\
G304.952+0.522 & MM & M/ND &\vline~ G11.956-0.177 & MM &  ND\\ % something weak at position
G304.933+0.546	& MM & ND &\vline~ G11.99-0.27 & M & M\\
G304.942+0.550	 & MM & ND &\vline~ G19.607-0.234 & MR & ND \\
G305.201+0.241	& MM &M &\vline~ G19.70-0.27A & M & ND \\ % looks liek a deteciton but noisy spectrum
G305.202+0.230	& MM & D/M &\vline~ G16.883-2.188 & MM & D/M\\  % baseline ripple, unreliable && too weak
G305.20+0.02 & R & ND&\vline~ G21.87+0.01 & MR & ND \\
G305.200+0.02	& M & ND &\vline~ G24.450+0.489 & MM & ND \\
G305.242+0.225	& MM & ND &\vline~ G23.420-0.235 & MM	&M \\ % &&too weak
G305.513+0.333	& MM & D/ND&\vline~ G23.319-0.298 & MM & S$^\gamma$\\ %  detection is solid, stable baseline, too weak
G305.533+0.360	& MM & ND&\vline~ G29.853-0.062& MM & ND	\\
\hline
\end{tabular}
\begin{scriptsize}
\begin{flushleft}
$^a$ Denotes the source class. See Section~\ref{sec:obs}.\\
$^b$ Code indicates whether the source is a: ND -- non-detection; M -- may be a detection, these sources are typically weak and/or the spectrum is too noisy to positively confirm a detection; D -- detection too weak compared with the noise; S -- Strong detection in which either (B) the hyperfines are blended  and can't be fit or  (A) has absorption features or ($^\gamma$) the spectrum is confused with multiple components and a fit will not converge.\\
$^c$ Two codes appear in this column only  if the \nhtwo\, transition is not in agreement with the \nhone, with the former following the latter and separated by a `/'.

\end{flushleft}
\end{scriptsize}
\end{table}

\subsection{Detection Rates}\label{sec:detection}

This Parkes ammonia (1,1) and (2,2) survey targeted a total of 244 sources of the 405 in the SIMBA sample of \citet{hill05}, i.e., 60 per cent. The breakdown of the different classes of source targeted and their detection rates in both \nhone\, and (2,2) is given in Table~\ref{tab:breakdown}. Of the 244 sources observed, 138 sources were detected (at a 3-$\sigma$ threshold) in \nhone\, (56 per cent) including two sources which had two velocity components in their spectra, and 102 in \nhtwo (42 per cent). Sources with a methanol maser or a radio continuum association (class M and R) have higher detection rates ($\sim$2/3) than class MM (52 per cent) and MR (50 per cent), though this distinction is only slight. Of the sources with an \nhone\, detection, 74 per cent were also detected in the (2,2) transition.  Class MR have the highest relative detection rate.

Compared with the number of sources targeted in each class, 60 per cent of the sample with active star formation (class M, MR and R); i.e., 59 of 96 sources, are detected in \nhone\, which decreases to 50 per cent for the (2,2) transition. In comparison, 50 per cent of the MM-only sources are detected in \nhone\, and only 36 per cent in \nhtwo.

\begin{table}
\caption{Summary of \nh\, detections}\label{tab:breakdown}
\begin{tabular}{@{}c@{}ccc@{}c}
\hline
Class & Sources  & \multicolumn{2}{c}{Good Fits} & \% of \nhone\,   \\
& Targeted & (1,1)  & (2,2) & with (2,2) \\
\hline
Total & 244 & 138 (56\%) &  102 (42\%) & 74\%\\ % 
MM & 148 & 79 (53\%) & 54 (36\%)& 68\%\\ 
M & 54 & 35 (65\%)  & 27  (50\%) & 75\%\\
MR & 24 & 12 (50\%) & 12 (50\%)& 100\%\\
R & 18 & 12 (67\%) & 9 (50\%) & 75\%\\
M+MR+R & 96 & 59 (61\%) & 48 (50\%) & 81\%\\
\hline
\end{tabular}
\end{table}

Of the remaining 108 sources not reported as detections in \nhone, 14 sources had obvious detections which could not be fit (see Table~\ref{tab:non_detections}). These tended to be sources quite close to the Galactic Centre and whose spectra displayed very broad lines with a combination of hyperfine blending, multiple lines and/or absorption features. A further 11 sources had detections that were too weak ($<$ 3-$\sigma$) in \nhone\, and a further 24 sources are possible detections though the noisy spectrum and very low signal-to-noise makes is difficult to determine whether there is a detection or not.

Three of the sources targeted have multiple velocity components in their spectra: G10.359-0.149; G23.268-0.257 and G332.646-0.647 which are visible in both the \nhone\, and (2,2) transition, though G23.268-0.257B did not pass our cuts described in section~\ref{sec:fitting}. For nine sources, the strength of the (2,2) transition is stronger than that of the (1,1), indicating that these sources are likely the hottest in the sample. It is not possible to determine an accurate column density or temperature for these sources, which are annotated by a $\dagger$ in Table~\ref{tab:nh3_11_22_fits}. These sources do not represent one particular class of source. As these transitions were observed simultaneously we can rule-out weather effects causing this.

\vspace{-0.5cm}
%__________________________________________________________________
\section{Results}

In this section the results pertaining to the parameters obtained from fitting the \nhone\, and (2,2) spectra (i.e., \delv, $\tau$, \vlsr, Flux density) are discussed, as well as the parameters derived from the fits: \tkin, \nhone\, column density (N$_{\nhone}$) and total \nh\, column density (N$_{\nh}^{TOT}$). Table~\ref{tab:averages} presents the mean and median values of each of these parameters.

\subsection{Linewidth \delv}\label{sec:linew}

The linewidth of the sample ranges from 1.1 to 7.1\,\kms\, for the \nhone\, transition and 1.3 to 9.2\,\kms\, for the \nhtwo\, transition. The mean linewidth of the \nhone\, data is 2.9\,\kms, with a standard deviation of 1.2\,\kms, whilst the (2,2) transition displays a slightly broader mean linewidth of 3.1\,\kms, with a standard deviation of 1.7\,\kms.

\citet{pillai06} found linewidths between 0.8 and 3\,\kms\, for their sample of nine infrared dark clouds (IRDCs), whilst for their sample of methanol maser selected sources, \citet{longmore07} find linewidths between 0.7 and 4.6\,\kms.  Both the linewidths of these authors and our sample, which is comprised of a cross-section of star-forming sources, display linewidths that are greater than those reported by \citet{jijina99} for their sample of low mass cores. \citet{pillai06} attribute the larger linewidths of their sample to velocity dispersions and turbulence. \citet{churchwell90b} find an average \nh\, linewidth of 3.1\,\kms\, for their sample of \uchii\, regions, whilst \citet{sridharan05} find a median linewidth of 1.5\,\kms\, for their sample of high mass starless cores and 1.9\,\kms\, for high mass protostellar objects. \citet{sridharan05} interpret the linewidth difference between the different types of source as an indication of evolutionary status, with more quiescent and less evolved cores having smaller linewidths.

\begin{figure}
\begin{center}
  \includegraphics[width=8cm, height=15cm]{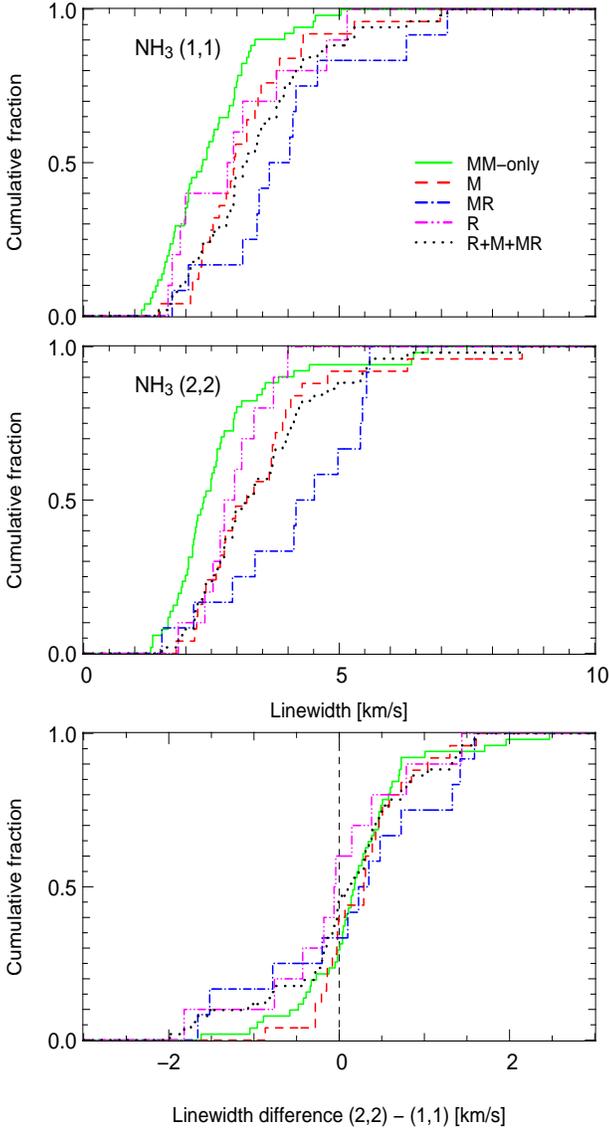}
\end{center}
\caption{Cumulative distribution of the \nhone\, (top) and \nhtwo\, (middle) linewidths of the sample. Key is as on plot. There is a slight offset between the (1,1) and (2,2) transitions. }\label{fig:cumul:line}

\end{figure}

The cumulative distribution plots of the \nhone\ and \nhtwo\, linewidth  (see Fig.~\ref{fig:cumul:line}), with respect to the class of source in the sample, confirms that the MM-only sources have the narrowest linewidths of the sample. Class R sources are confined to a small region of the linewidth parameter space which may simply reflect the small number of sources in this sample (see Table~\ref{tab:breakdown}). For both the \nhone\, and (2,2) distributions, the maser sample traces that of the sample with active star formation (class M+MR+R) indicating that masers are ubiquitous tracers of massive star formation. Class MR sources have the broadest linewidths of the sample, given that they have two indicators of active star formation they could potentially be warmer and/or more evolved. 

To test the hypothesis that two, or more, classes of source are drawn from the sample parent population, Kolmogorov-Smirnov (KS) tests were performed. For both the (1,1) and (2,2) transition, the KS-tests for the individual classes were inconclusive which is likely due to the low number of sources in the individual samples of active sources (M, MR and R).  When comparing the MM-only sample (class MM) with the star formation sample (class M+MR+R), the statistics become significant enough to perform robust KS tests.  The null hypothesis, that the samples are drawn from the same population, can clearly be rejected when comparing the MM-only sample (class MM) with the star formation sample (class M+MR+R) -  see full red and dotted black line in Fig.~\ref{fig:cumul:line}. This results holds for both the (1,1) and (2,2) transition, with probabilities smaller than  10$^{-3}$ and 7$\times$10$^{-5}$, respectively.

As the \nhone\, and (2,2) transition are expected to be emitted from the same gas, we would then expect that they display similar linewidths. The top two panels of Figure~\ref{fig:cumul:line} suggest that there is a slight increase in linewidth with an increase in the \nh\, transition. Note that the \nhone\, plot includes all sources, whilst the \nhtwo\, plot includes sources with a Rel. Group equal to 1 or 2, as Group 3 sources have only upper limits for this transition. This result is in agreement with \citet{pillai06} who also found that for some of their IRDCs the (2,2) linewidths were slightly larger than the (1,1) linewidths, which they interpreted to mean that each transition was not exactly tracing the same gas. Broader \nhtwo\, linewidths, with respect to the \nhone\, linewidth, could be interpreted as internal heating, but further (mapping) observations of these sources are necessary to examine this.

The bottom panel of Figure~\ref{fig:cumul:line} is a plot of the difference between the \nhone\, and (2,2) linewidth. This panel places the top and middle panel in context, indicating that approximately half of the sources have broader (2,2) linewidths and the other half have less broad line widths compared with the \nhone\, transition. This panel also clearly shows that each of the classes of source have comparable (2,2) versus (1,1) linewidths. If we refer to the sources for which we have robust detections (i.e. a Rel. Group equal to 1 or 2), the (2,2) linewidths can vary from -2.0 to 2.5\,\kms\, broader than the (1,1) transition with a median and mean of 0.0\,\kms.

For the 45 per cent of sources with \nhone\, linewidths broader than the (2,2) transition, the (2,2) linewidth is 50 to 99 per cent that of the width of the (1,1) transition, with a median of 90 per cent of the (1,1) linewidth. The small difference in linewidth ensures that we can use the two transitions to determine the temperature of the cores. 

Thermal linewidths range from 0.1 to 0.4\,\kms\, with a median of 0.2\,\kms, as calculated from the \nh\, derived kinetic temperature. The turbulence is clearly dominating for all sources, including the MM-only where the linewidth is the smallest.

\subsection{Optical Depth $\tau$}

The optical depth of the sample ranges from 0.1 to 3.8 for the \nhone\, transition and from 0.1 to 9.4 for the (2,2) transition. We caution however that this lower limit is simply the minimum value for optical depth as determined from CLASS, and should simply be interpreted as optically-thin emission. The median optical depth of the \nhone\, is 1.4 compared with 0.1 for the (2,2) transition. These optical depth values are consistent with \citet{longmore07} who found that the majority of their maser sources have optical depths between 0.3 and 5. There is no obvious difference in the optical depth between the different classes of source for either of the \nh\, transitions. Only 22 per cent of sources observed in both (1,1) and (2,2) have greater optical depths for the (2,2) transition. 

\subsection{\vlsr}

The \vlsr\, of the sample ranges from -87.8 to 113\kms.  There are no trends between the source \vlsr\, and type of source in the sample.

\subsection{Flux Density}

The \nhone\, flux density of the sample ranges from 0.3 to 11.60\,Jy/beam, both of which are MM-only sources, with a median of 2.4\,Jy/beam. The different classes of source in the sample display a similar range of flux. Class M, MM and MR have similar median flux density values of 2.5\,Jy/beam, whilst class R have a median flux of 1.6\,Jy/beam, though the sample of active star formation sources (class M, MR and R) has a median value of 2.4\,Jy/beam. For nine sources, the flux density of the (2,2) transition is greater than that of the (1,1) -- see section~\ref{sec:detection}.

For the \nhtwo\, transition the flux density of the sample ranges from 0.3 to 6.4\,Jy/beam, with a median of 1.2\,Jy/beam. Each of the four classes of source display similar flux ranges for this transition. The median \nhtwo\, flux density of each class suggests that the MM-only sources are not as bright as the other classes of source. We caution however that this difference is small. The \nhtwo\, flux density of the MM-only cores is 75 per cent that of the sample with active star formation.

Cumulative distributions and Komolgorov-Smirnov (KS) tests indicate that there is little difference between each of the classes in terms of their flux density.

The peak line intensity of the \nhtwo\, transition for our sources is on average 40 per cent that of the (1,1) transition (excluding the nine sources which have greater (2,2) flux densities). This ratio is in excellent agreement with \citet{pillai06} who also find that for their sample of IRDCs, the peak intensity of the (2,2) is 40 per cent that of the (1,1).

\subsection{Kinetic Temperature \tkin}\label{sec:tkin}

The kinetic temperature of the sample ranges from 9.0 to 98.0\,K, with a median of 20\,K. Despite such a large temperature range, only 22 per cent of sources have temperatures greater than 30\,K, whilst only 6 per cent have temperatures in excess of 50\,K. 78 per cent of sources are cooler than 30\,K. As discussed in section~\ref{sec:fitting} and \ref{sec:dis:sed}, the lower transition (1,1) and (2,2) \nh\, data are useful as a thermometer when the temperature is less than 30\,K \citep{danby88}. When the temperature is greater than this, these transitions can not support accurate temperature determinations. \citet{tafalla04} are more conservative and suggest that \nhone\, and (2,2) actually poorly constrain temperatures greater than 20K. However, our method of modelling, i.e., Bayesian inference, factors in the larger error bars for ill constrained temperatures, when deriving masses from SEDs - see section \ref{sec:remodel}.

The MM-only sources are cooler on average than each of the individual classes comprising maser and/or radio continuum associations as well as the sample of active star formation (class M+MR+R) - see Table~\ref{tab:averages}. The cumulative distribution plot of the temperature is discussed in detail in section \ref{sec:sed}.

\citet{pillai06} found that the temperature of their IRDC sample range from 11--17\,K, whilst \citet{churchwell90b} find approximately half of their \uchii\, sample to range between 15 and 25\,K and the other half to have temperatures in excess of 25\,K. 

\subsection{Total Column Density}

\begin{figure}
\begin{center}
 \includegraphics[width=0.85\hsize]{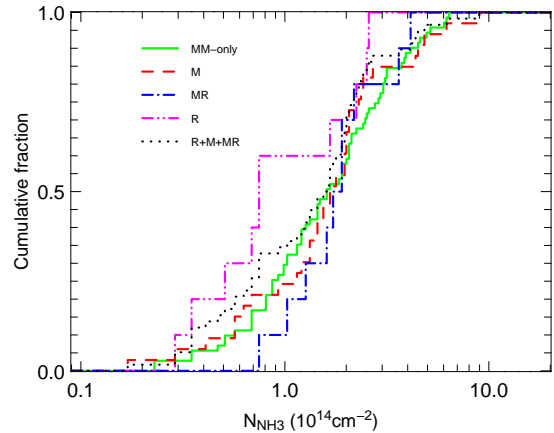}
\caption{Cumulative distribution of the column density of the different classes in the sample. Key is on the plot. \label{fig:colden}}
\end{center}
\end{figure}

The total column density of the sources, derived from both the \nhone\, and (2,2) transitions, ranges from 0.3 to 15.7\,$\times$10$^{14}$~cm$^{-2}$, with a median column density of 2.8\,$\times$10$^{14}$~cm$^{-2}$. Figure~\ref{fig:colden} presents the cumulative distribution plot of the column density. The radio sample (class R) has the lowest column density on average compared with the other classes -- see Table~\ref{tab:averages}. 

\subsection{Correlations}

Cross-correlating each of the parameters discussed in the aforementioned sections reveals very few trends and thus relationships between them. There is a slight correlation between the flux and the optical depth of a sources, with an increase in flux translating to an increase in the optical depth of the source.

Both \citet{longmore08} and \citet{pillai06} found that the \nhone\, linewidth and kinetic temperature of a sources are correlated, with an increase in one leading to an increase in the other. \citet{longmore08} found that when they compared these two parameters, the \nh-only cores of their sample -- those without methanol maser associations -- were confined to a small and distinct region in the plot, which was clearly separate from the parameter space occupied by the methanol maser sources. The \nh-only sources were cooler than the maser sources, yet of comparable linewidth.

As mentioned in section \ref{sec:linew}, \citet{pillai06} found that IRDCs in their sample have greater linewidths than the low-mass stars in the sample of \citet{jijina99}. Additionally, they find that their IRDCs are colder relative to the sample of \citet{jijina99} and the \uchii\, sample of \citet{churchwell90b}. A comparison of their IRDCs with more evolved examples of massive star formation (e.g. \uchii\, regions and high mass protostellar objects e.g. \citealp{beuther02}) reveals the IRDCs to be cooler than these sources and to have narrower linewidths on average.

We do not see this correlation in our own data, i.e., the MM-only sources are not distinct from the other star-formation classes with respect to their temperature. Instead, we find that the MM-only sample has comparable temperatures to sources with a methanol maser and/or radio continuum sources, coupled with smaller linewidths. The eight dark clouds for which we have \nhone\, and (2,2) observations are not confined to cooler temperatures or smaller linewidths as per \citet{pillai06}, nor is there a trend between their \nhone\, linewidth and \tkin. The lack of correlation between the \nhone\, and \tkin\, of our sample may be attributed to two factors. Firstly, our data are likely subject to beam dilution with low-resolution (58 arcsec), compared with the 40 arcsec resolution of \citet{pillai06} and the interferometric observations  ($<$\,11 arcsec)  of \citet{longmore08}. Additionally the combination of \nhone\, and (2,2) provides reliable temperatures up to $\sim$\,30\,K, so we are insensitive to the greater temperature range coverage of \citet{longmore08} who used higher transition ammonia data to determine their temperatures. 

While \citet{longmore06} were able to probe warmer regions better than us, due to higher transition \nh\, data which they also measured, when it comes to the cold sources of particular interest here, this is not an issue. Both studies are sensitive to probing the temperature in the coldest gas, where only the (1,1) and (2,2) lines are significantly excited.

\begin{table*}
\caption{Mean and median as derived from the ammonia spectra for each of the source classes.  For a breakdown of the number of sources within each class, refer to Table  \ref{tab:breakdown}.}
\label{tab:averages}
\begin{tabular}{@{}llcccccc@{}}
\hline 
& &\multicolumn{4}{c}{Class} &whole &all except \\
Parameter& & MM-only & maser & maser+radio & radio& sample &  MM-only \\
& & (MM) & (M) & (MR) & (R) &&(M+MR+R)\\
\hline
Flux Den.$^{1,1}$  & mean & 3.3 & 3.0 & 2.9 & 2.2 & 3.1 & 2.8 \\ 
   (Jy/beam)          & median &2.6& 2.6 & 2.5 & 1.6 & 2.5 & 2.4 \\
\hline
$\Delta$V$^{1,1}$ & mean & 2.6 & 3.1 & 3.4 & 3.0 & 2.8 & 3.1\\
(km.s$^{-1}$)     & median& 2.3 & 2.9 & 3.5 & 2.4 & 2.7 & 3.0\\ 
\hline 
Flx Den.$^{2,2}$ & mean & 1.3 & 1.5 & 1.3 & 1.0 & 1.3 & 1.4 \\
(Jy/beam)           &median & 1.1 & 1.5 & 1.2& 0.9 & 1.2 & 1.3 \\
\hline
$\Delta$V$^{2,2}$  & mean & 2.6 & 3.5& 3.9& 2.7& 3.0& 3.5\\ 
(km.s$^{-1}$)     & median & 2.3 & 3.3 & 4.1 & 2.7& 2.6& 3.2\\
\hline
 T$_{\rm kin}$ & mean & 22.4 & 29.8& 32.2& 28.5& 25.8 & 30.2\\ 
(K)  & median & 19.0 & 25.0& 24.0& 23.5& 20.0& 25.0 \\
\hline
 N$_{NH3}^{TOT}$ & mean & 3.4 & 3.7& 3.5& 2.2& 3.4& 3.4\\ 
(10$^{14}$cm$^{-2}$) & median & 2.7& 2.9& 3.2& 1.3& 2.8& 2.9\\
\hline
\end{tabular}
\end{table*}

%__________________________________________________________________
\section{Analysis}

\subsection{Previous SED modelling}\label{sec:sed}

\citet{hill09} performed spectral energy distribution modelling of $\sim$\,180 sources of the SIMBA sample (see section~\ref{sec:intro}), using the Bayesian inference method of fitting \citep[e.g.][]{pinte08}.  This method of fitting considers the potential correlations between parameters to produce quantitative estimates of the range of validity of key parameters (temperature, mass, luminosity) extracted from SED fitting. As the modelling procedure is outlined in detail in section 3.1 of \citet{hill09} we simply summarise the procedure here. 

The sources were modelled according to a two-component model which denotes a central warm core surrounded by a cold dust envelope. The hot component of this model is assumed to radiate as a blackbody whilst the cold component accounts for optically thin emission from the dust \citep[see Equation 1,][]{hill09}. The sources were then fit for four free parameters:  $R_\mathrm{hot}$, $T_\mathrm{hot}$, $T_\mathrm{cold}$ and $M_\mathrm{cold}$. Those sources without mid-infrared emission (i.e, a hot component) were fit for the cold component only and thus two free parameters: $T_\mathrm{cold}$ and $M_\mathrm{cold}$.  

It was clear from SED modelling \citep{hill09} that an absence of far-infrared data, where the peak of the dust emission lies, hinders accurate determinations of the source temperature. Additionally as the mass and luminosity of a source are highly correlated with that of the temperature, assessing the evolutionary  status of a source from SED fitting alone is difficult. Greater observational constraints were necessary in order to facilitate further SED fitting and analysis.
 
\subsection{SED modelling revisited}

\subsubsection{Comparison of temperature derivation methods}

These ammonia data provide an independent, and more accurate, determination of the source temperature. Following the assumption that the kinetic temperature is equivalent to the dust temperature, it is then possible to revisit our previous SED modelling. \citet{li03} showed that the gas temperature of their sources i.e., the kinetic temperature as derived from \nh\, observations, was within a few K of the dust temperature. According to their observations, \tkin\, and T$_{\rm dust}$ are expected to be similar in cold regions. \citet{schnee07} also find excellent agreement between the dust and gas temperature of their star less core in Taurus. Contrary to this, both \citet{molinari96} and \citet{sridharan02} find discrepancies when comparing dust temperatures and kinetic temperatures derived from \nh. However, both these authors derived their dust temperatures from \IRAS\, data which is subject to poor resolution and the emission is likely optically thick.  When the dust is highly optically thick large differences between the gas and dust temperatures can occur \citep{kruegel84}. \citet{molinari96} themselves caution that IRAS fluxes alone are insufficient for proper estimates of the dust temperature.

\begin{figure}
\begin{center}
  \includegraphics[width=8cm, height=7cm]{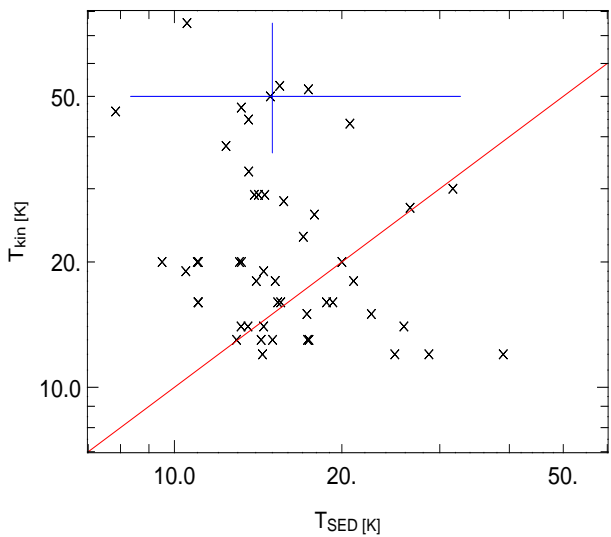}
\caption{Comparison of the temperature as derived from SED modelling with that derived from the \nh\, observations. The red line indicates where these two parameters would be equal. The error bars are indicated by the blue lines. \label{fig:T_comparison}}
\end{center}
\end{figure}

Of the $\sim$~180 sources that were originally SED modelled \citep{hill09}, 82 were also observed with ammonia, though 30 of these sources only have upper limits to their temperature. Figure~\ref{fig:T_comparison} compares the kinetic temperature (\tkin) of each source, as derived from the \nh\, observations, with the dust temperature as estimated from our previous SED modelling (T$_{\rm SED}$). This figure contains the 52 source with reliable continuum and ammonia observations for estimating both temperatures (Rel. Group 1 and 2, in Table~\ref{tab:nh3_derivs}), thus allowing a statistical comparison of both methods. The red line on the plot indicates where T$_{\rm SED}$ and \tkin\, are equal. 

Figure~\ref{fig:T_comparison} shows that there are no systematic biases introduced in either of the methods used to determine the temperature estimates.  If SED fitting was biasing the resultant temperature of a core, then these data would all be above, or below, the red line. The SED method, although not strongly constraining the temperature, appears a robust method for obtaining a first order unbiased estimate of the core temperature.  In this section, we perform the SED modelling again, this time using the kinetic temperature as derived from \nh. 

\subsubsection{SED modelling with \nh}\label{sec:remodel}

SED modelling was performed using the Bayesian inference method, following the procedure summarised above (section~\ref{sec:sed}) and outlined in \citet[][section 3]{hill09}.
Rather than fitting for the temperature ($T_\mathrm{cold}$), we fixed the dust temperature to the kinetic temperature, as derived from \nh, and fit solely for the mass of the source (as well as the hot parameters, for those sources with mid-infrared data), assuming the same dust properties used previously. As per \citet{hill09} we have assumed the near distance for all sources with a distance ambiguity. 

The probability distributions of the temperature were set to Gaussian distributions defined by the value of, and uncertainties with, the \nh\, temperature (\tkin) - see Table~\ref{tab:nh3_11_22_fits}. For those sources where only an upper limit value to the temperature has been determined (Rel. Group 3) we assume a uniform probability distribution between 2.73\,K and the upper limit. The estimated range of validity for the mass and the luminosity derived from this method then not only accounts for the uncertainty on the temperature but also considers the correlations between parameters.

\begin{figure*}
  \includegraphics[width=0.85\hsize]{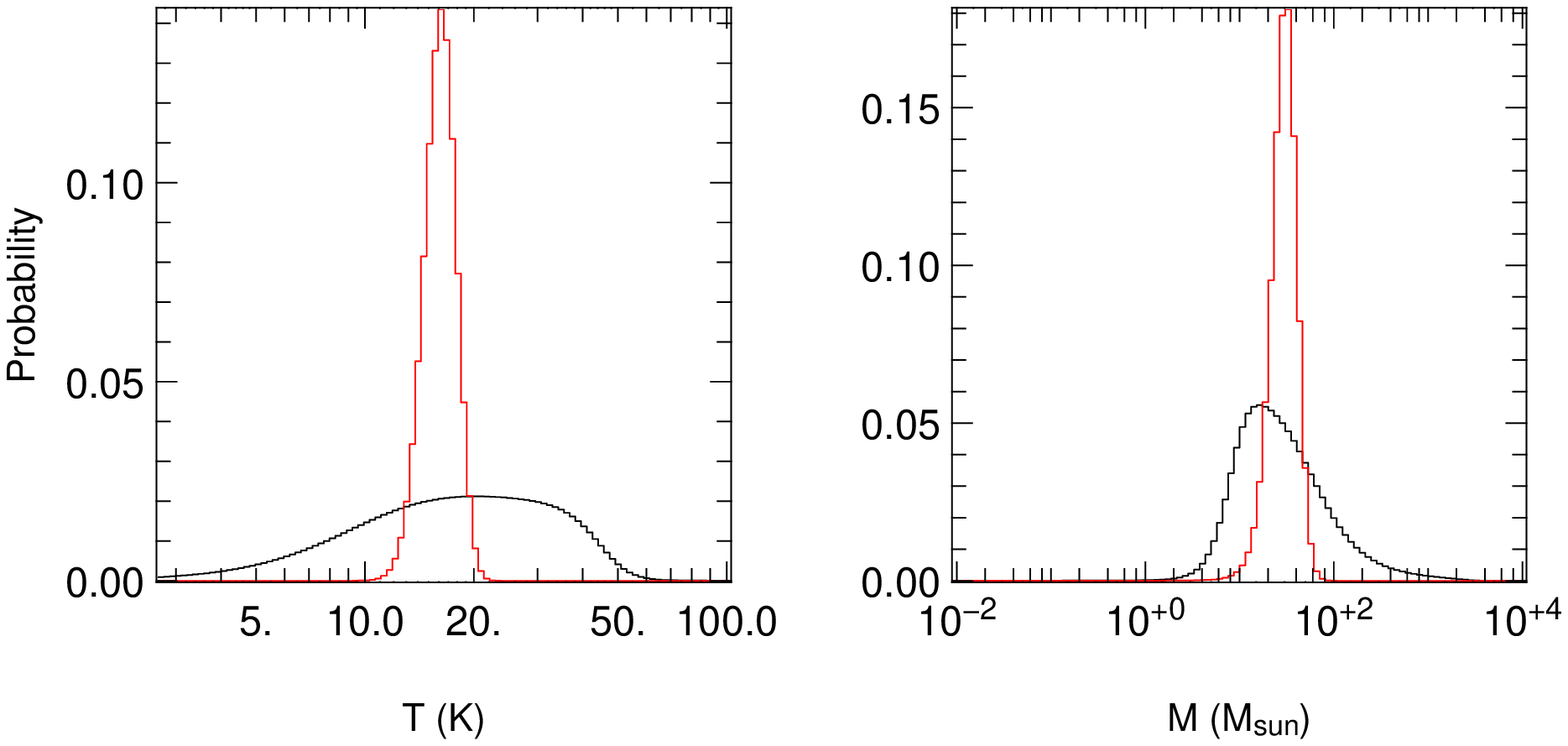}
\caption{The probability curves of one source \citep[G10.288--0.127, corresponding to the bottom row of Fig. 1][]{hill09}  comparing the two SED methods. The black line indicates the probability curve corresponding to a SED fit temperature \citep{hill09} whilst the red probability curve is indicative of the probability when the temperature is fixed using the ammonia derived \tkin. As can be seen, constraints on the temperature translate as tighter probability curves for parameters resulting from SED modelling, i.e., mass.
\label{fig:proba}}
\end{figure*}

Figure~\ref{fig:proba} is a plot of the probability curves for the temperature and mass of a source, comparing our previous SED fitting method \citep{hill09}, with that of the SED fitting performed in this paper using \nh\, derived temperatures.  Figure~\ref{fig:proba} illustrates how kinetic temperatures allow tighter constraints on the mass of a source, as is expected. This source,  G10.288--0.127 a MM-only source with no mid-infrared association which was consequently fit with a single cold component, corresponds to the bottom row of Fig.~1 of \citet{hill09}. This source was chosen to illustrate how a source with limited sampling of the SED can be well constrained by known or constrained temperatures. It is clear that the \nh\, data (red curve) provides greater constraints on the temperature (left-hand plot), especially for those sources with few data points, compared with deriving the temperature from SED fitting (black curve). The range of validity for the temperature is now constrained to a smaller portion of the parameter space.   It is clear that accurate temperatures are necessary to derive accurate sources masses.

\subsection{SED results}\label{sec:dis:sed}

\begin{table*}
\caption{Parameters resulting from SED modelling of 52 SIMBA sources with \nh\, temperatures reported in Table \ref{tab:nh3_derivs}. For each of the temperature, mass and luminosity, the range corresponding to a 1-sigma (68 per cent) probability of occurrence is presented, with the $_{min}$ and $_{max}$ values representing the lower and upper values of this range, respectively.\label{tab:sed_values}}
\begin{tabular}{@{}lccccrrrr@{}}
\hline 
 & Ident & Fit & \multicolumn{2}{c}{Temperature} & \multicolumn{2}{c}{Mass}& \multicolumn{2}{c}{Luminosity} \\
 Source Name & tracer & Type  & T$_{\rm cold}$$_{_{min}}$ &  T$_{\rm cold}$$_{_{max}}$ &  M$_{min}$ & M$_{max}$ & L$_{min}$ & L$_{max}$\\
    &        & &  K   & K   &  \mstar     & \mstar & \lum & \lum\\
\hline
 G305.833-0.196  &  mm   &  SINGLE   &  26  &  28  &  7.6E+01  &  1.0E+02  &  3.7E+03   &  1.6E+04 \\
 G323.74-0.3  &  m   &  SED   &  15  &  37  &  2.0E+02  &  2.8E+03  &  5.3E+03   &  1.0E+05 \\
 G0.32-0.20  &  mr   &  SINGLE$^\alpha$   &  17  &  31  &  3.3E+03  &  8.7E+03  &  3.4E+04   &  6.7E+05 \\
 G1.105-0.098  &  mm   &  SINGLE$^\alpha$   &  28  &  47  &  5.3E+02  &  1.4E+03  &  1.0E+05   &  9.8E+05 \\
 G1.13-0.11  &  r   &  SED$^\alpha$   &  34  &  47  &  3.1E+02  &  3.8E+03  &  4.6E+05   &  1.4E+06 \\
 G0.549-0.868  &  mm   &  SINGLE$^\alpha$   &  15  &  24  &  2.5E+01  &  5.0E+01  &  6.1E+01   &  5.3E+03 \\
 G0.600-0.871  &  mm   &  SINGLE$^\alpha$   &  11  &  42  &  1.9E+01  &  1.3E+02  &  8.9E+01   &  7.2E+04 \\
 G2.54+0.20  &  m   &  SINGLE   &  17  &  26  &  2.0E+02  &  4.6E+02  &  5.7E+02   &  2.4E+04 \\
 G5.504-0.246  &  mm   &  SINGLE$^\alpha$   &  10  &  19  &  1.9E+03  &  5.0E+03  &  8.3E+02   &  3.4E+04 \\
 G5.90-0.42  &  m   &  SED$^\alpha$   &  15  &  29  &  5.3E+02  &  1.9E+03  &  5.3E+03   &  7.2E+04 \\
 G8.111+0.257  &  mm   &  SINGLE$^\alpha$   &  21  &  83  &  6.1E+00  &  3.3E+01  &  5.7E+02   &  6.7E+05 \\
 G8.127+0.255  &  mm   &  SINGLE$^\alpha$   &  14  &  32  &  5.7E+01  &  2.0E+02  &  2.7E+02   &  5.0E+04 \\
 G8.138+0.246  &  mm   &  SINGLE$^\alpha$   &  15  &  31  &  1.1E+02  &  4.0E+02  &  5.7E+02   &  7.2E+04 \\
 G8.13+0.22  &  mr   &  SED$^\alpha$   &  14  &  20  &  9.3E+02  &  2.5E+03  &  3.7E+03   &  3.4E+04 \\
 G5.962-1.128  &  mm   &  SINGLE$^\alpha$   &   8  &  31  &  8.1E+00  &  7.6E+01  &  6.6E+00   &  1.6E+04 \\
 G5.975-1.146  &  mm   &  SINGLE$^\alpha$   &  17  &  50  &  7.1E+00  &  2.5E+01  &  1.9E+02   &  7.2E+04 \\
 G9.63+0.19  &  mr   &  SED$^\alpha$   &  16  &  36  &  2.0E+02  &  1.1E+03  &  2.5E+03   &  5.0E+04 \\
 G8.68-0.36  &  mr   &  SINGLE$^\alpha$   &  11  &  16  &  6.6E+03  &  1.1E+04  &  3.7E+03   &  3.4E+04 \\
 G8.686-0.366  &  m   &  SINGLE$^\alpha$   &  16  &  23  &  1.1E+03  &  2.2E+03  &  5.3E+03   &  7.2E+04 \\
 G10.287-0.110  &  mm   &  SINGLE$^\alpha$   &  12  &  35  &  5.7E+01  &  3.1E+02  &  2.7E+02   &  7.2E+04 \\
 G10.284-0.126  &  m   &  SED$^\alpha$   &  18  &  45  &  4.3E+01  &  2.3E+02  &  1.2E+03   &  3.4E+04 \\
 G10.288-0.127  &  mm   &  SINGLE$^\alpha$   &  19  &  32  &  2.8E+01  &  6.6E+01  &  3.9E+02   &  1.6E+04 \\
 G10.29-0.14  &  mr   &  SED$^\alpha$   &  16  &  25  &  2.7E+02  &  7.1E+02  &  3.7E+03   &  2.4E+04 \\
 G10.343-0.142  &  m   &  SINGLE$^\alpha$   &  15  &  35  &  5.0E+01  &  1.5E+02  &  2.7E+02   &  5.0E+04 \\
 G10.63-0.33B  &  mm   &  SINGLE$^\alpha$   &  16  &  62  &  1.7E+02  &  1.2E+03  &  1.1E+04   &  4.3E+06 \\
 G10.62-0.33  &  m   &  SED$^\alpha$   &  14  &  28  &  9.3E+02  &  3.8E+03  &  5.3E+03   &  7.2E+04 \\
 G9.88-0.75  &  r   &  SED$^\alpha$   &  13  &  20  &  1.2E+03  &  3.3E+03  &  2.5E+03   &  1.6E+04 \\
 G10.62-0.38  &  mr   &  SINGLE$^\alpha$   &  12  &  32  &  6.6E+03  &  3.5E+04  &  2.4E+04   &  2.1E+06 \\
 G11.11-0.34  &  r   &  SED$^\alpha$   &  13  &  28  &  8.1E+02  &  3.8E+03  &  3.7E+03   &  7.2E+04 \\
 G11.117-0.413  &  mm   &  SINGLE$^\alpha$   &  11  &  14  &  4.6E+02  &  8.1E+02  &  8.9E+01   &  3.7E+03 \\
 G12.88+0.48  &  m   &  SINGLE$^\alpha$   &  14  &  28  &  9.3E+02  &  3.3E+03  &  1.7E+03   &  1.0E+05 \\
 G12.914+0.493  &  mm   &  SINGLE$^\alpha$   &   9  &  36  &  5.7E+01  &  4.6E+02  &  8.9E+01   &  7.2E+04 \\
 G11.903-0.140  &  mr   &  SINGLE$^\alpha$   &   9  &  17  &  5.3E+02  &  1.6E+03  &  1.3E+02   &  1.1E+04 \\
 G12.18-0.12A  &  m   &  SINGLE$^\alpha$   &  14  &  19  &  2.5E+03  &  4.3E+03  &  3.7E+03   &  3.4E+04 \\
 G12.216-0.119  &  mm   &  SINGLE$^\alpha$   &  16  &  31  &  2.2E+03  &  5.7E+03  &  2.4E+04   &  6.7E+05 \\
 G12.43-0.05  &  r   &  SED   &  16  &  30  &  2.2E+03  &  8.7E+03  &  1.1E+04   &  3.2E+05 \\
 G12.68-0.18  &  m   &  SED$^\alpha$   &  11  &  17  &  1.9E+03  &  4.3E+03  &  2.5E+03   &  1.1E+04 \\
 G11.94-0.62B  &  mm   &  SINGLE$^\alpha$   &  11  &  13  &  1.6E+03  &  2.5E+03  &  5.7E+02   &  5.3E+03 \\
 G11.93-0.61  &  mr   &  SED$^\alpha$   &  13  &  15  &  1.6E+03  &  3.3E+03  &  1.7E+03   &  7.7E+03 \\
 G12.90-0.25B  &  mm   &  SINGLE$^\alpha$   &  11  &  12  &  7.1E+02  &  1.1E+03  &  1.9E+02   &  3.7E+03 \\
 G13.87+0.28  &  m   &  SED   &  10  &  39  &  4.6E+02  &  6.6E+03  &  7.7E+03   &  3.2E+05 \\
 G12.859-0.272  &  mm   &  SED$^\alpha$   &  15  &  40  &  1.3E+02  &  8.1E+02  &  1.2E+03   &  3.4E+04 \\
 G12.90-0.26  &  m   &  SED$^\alpha$   &  14  &  19  &  1.6E+03  &  3.8E+03  &  5.3E+03   &  2.4E+04 \\
 G14.60+0.01  &  mr   &  SED$^\alpha$   &  11  &  31  &  1.3E+02  &  9.3E+02  &  3.9E+02   &  1.1E+04 \\
 G10.84-2.59  &  r   &  SINGLE$^\alpha$   &  12  &  30  &  1.3E+02  &  6.1E+02  &  1.9E+02   &  3.4E+04 \\
 G16.58-0.05  &  m   &  SED$^\alpha$   &  14  &  21  &  8.1E+02  &  2.2E+03  &  2.5E+03   &  1.6E+04 \\
 G18.30-0.39  &  r   &  SED   &  14  &  36  &  1.7E+02  &  1.4E+03  &  5.3E+03   &  7.2E+04 \\
 G19.61-0.1  &  m   &  SED$^\alpha$   &  13  &  36  &  1.3E+02  &  1.1E+03  &  5.7E+02   &  5.0E+04 \\
 G16.86-2.15  &  m   &  SED$^\alpha$   &  11  &  17  &  8.1E+02  &  2.2E+03  &  8.3E+02   &  3.7E+03 \\
 G23.71+0.17  &  r   &  SED$^\alpha$   &  12  &  20  &  1.6E+03  &  5.0E+03  &  7.7E+03   &  7.2E+04 \\
 G29.96-0.02B  &  mr   &  SED$^\alpha$   &  13  &  31  &  2.2E+03  &  1.0E+04  &  1.1E+04   &  4.6E+05 \\
 G29.912-0.045  &  mm   &  SINGLE$^\alpha$   &  14  &  20  &  1.9E+03  &  3.8E+03  &  3.7E+03   &  5.0E+04 \\
\hline
\end{tabular}
\begin{flushleft}
$^\alpha$ denotes those sources that were fit with \IRAS\, upper limits - see \citet{hill09}.
\end{flushleft}
\end{table*}
% 13 16 58.38 & -62 55 25.2 &
% 15 31 44.50 & -56 30 51.0 &
% 17 47 09.71 & -28 46 08.0 &
% 17 48 36.41 & -28 02 31.0 &
% 17 48 42.46 & -28 01 35.0 &
% 17 50 18.77 & -28 53 19.0 &
% 17 50 26.07 & -28 52 31.0 &
% 17 50 46.50 & -26 39 44.0 &
% 17 59 07.53 & -24 19 19.0 &
% 18 00 40.90 & -24 04 12.0 &
% 18 02 49.31 & -21 48 34.0 &
% 18 02 52.76 & -21 47 54.0 &
% 18 02 56.21 & -21 47 38.0 &
% 18 03 01.95 & -21 48 02.0 &
% 18 03 29.19 & -24 21 49.0 &
% 18 03 33.88 & -24 21 41.0 &
% 18 06 14.80 & -20 31 29.0 &
% 18 06 18.91 & -21 37 21.0 &
% 18 06 23.49 & -21 36 57.0 &
% 18 08 45.85 & -20 05 42.0 &
% 18 08 49.25 & -20 05 58.0 &
% 18 08 52.66 & -20 05 58.0 &
% 18 08 56.07 & -20 05 50.0 &
% 18 09 00.04 & -20 03 34.0 &
% 18 10 15.59 & -19 54 45.0 &
% 18 10 18.42 & -19 54 29.0 &
% 18 10 19.00 & -20 45 25.0 &
% 18 10 28.77 & -19 55 48.0 &
% 18 11 31.80 & -19 30 44.0 &
% 18 11 35.76 & -19 30 44.0 &
% 18 11 51.40 & -17 31 30.0 &
% 18 11 53.64 & -17 30 02.0 &
% 18 12 11.11 & -18 41 30.0 &
% 18 12 43.25 & -18 25 09.0 &
% 18 12 44.37 & -18 24 21.0 &
% 18 12 54.72 & -18 11 04.0 &
% 18 13 54.14 & -18 01 41.0 &
% 18 13 58.08 & -18 54 14.0 &
% 18 14 00.90 & -18 53 18.0 &
% 18 14 33.90 & -17 51 44.0 &
% 18 14 35.54 & -16 45 36.0 &
% 18 14 36.13 & -17 54 56.0 &
% 18 14 38.94 & -17 51 52.0 &
% 18 17 02.17 & -16 14 28.0 &
% 18 19 12.03 & -20 47 23.0 &
% 18 21 09.10 & -14 31 40.0 &
% 18 25 41.65 & -13 10 16.0 &
% 18 27 16.34 & -11 53 51.0 &
% 18 29 24.20 & -15 16 06.0 &
% 18 33 53.06 & -08 07 23.0 &
% 18 46 03.97 & -02 39 25.0 &
% 18 46 05.04 & -02 42 29.0 &

\begin{figure*}
  \includegraphics[angle=270,width=0.95\hsize]{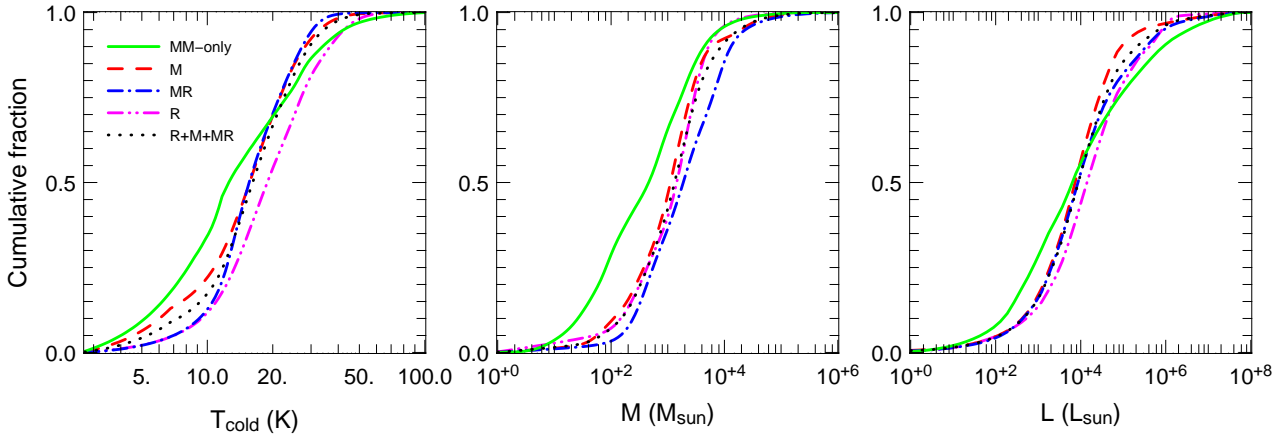}
\caption{Cumulative distributions for the SIMBA sources that were re-modelled with SEDs using their \nh\, temperature. The key on the left plot indicates the different classes of source in the sample. left: distribution of temperature. middle: distribution of mass. right: distribution of luminosity. 
 \label{fig:sedcumul}}
\end{figure*}

Table~\ref{tab:sed_values} presents the range of validity for the temperature,mass and luminosity of each of the 52 sources which were re-modelled using their \nh\, temperature in the SED fits. The cumulative distribution plots of the temperature, mass and luminosity are presented in Figure~\ref{fig:sedcumul}. Note that all sources, regardless of their reliability flag (Table~\ref{tab:nh3_derivs}) were included in this Figure.

An interesting result evident from the probability distributions of all of the sources re-modelled is the complementarity of the SED fitting and \nh\, observations in terms of parameter determinations. Even when the uncertainty on the kinetic temperature derived from \nh\, is large, the constraints from the SED fitting and the \nh\, analysis are very complementary. SED modelling provides good constraints on the upper limit of the temperature when the ammonia data are ill-constrained due to the absence of higher transitions (see section~\ref{sec:fitting}). On the other hand, \nh\, provides good constraints on the lower limit to the temperature when the ambiguities from SED modelling become very large.

Comparison of the temperature-fit probability curves \citep{hill09} with those of the probability curves produced here, indicates that, in most instances, the mass probability peaks in the same place for both methods. In all instances, the probability curve of the mass for each individual source is more tightly constrained using the \nh\, temperatures, as is demonstrated in Fig.~\ref{fig:proba}. 

The cumulative distribution plot of the temperature (Fig.~\ref{fig:sedcumul}, left)  suggests that class MM sources are slightly cooler than the other classes of source. This difference reflects the fact that 9 of the 11 sources in Rel. Group 3, and thus sources with upper limits, are MM-only sources which effectively add a `bump' to the MM-only curve at low temperatures. However, the difference between the MM-only sources and those with signatures of massive star formation is not statistically significant, with a KS test probability of 0.4. If we do not include the sources in Rel. Group 3, the distribution of the temperature shows little distinction between the different classes of source in the sample for this parameter with a KS probability of 0.98.

The cumulative distribution plot of the mass (Fig.~\ref{fig:sedcumul}, middle) indicates that the MM-only sources are the least massive of the sample, whilst class MR are the most massive. The difference in the mass is more significant than for the temperature, though the KS test, with a probability of 0.08, does not allow us to rule on the null hypothesis that MM-only sources are drawn from the same population as that star formation sources. This is likely a result of the relatively small number of sources modelled.

The cumulative distribution plot of the luminosity (Fig.~\ref{fig:sedcumul}, right) reveals the MM-only sources to be marginally less luminous than the other classes of source,  which is not surprising given that the mass and luminosity are correlated and the temperature of the sources is slightly smaller.

\subsection{Luminosity vs Mass Diagram}

As per \citet{hill09}, we have drawn a luminosity vs. mass (hereafter $M-L$ diagram) diagram for the sources which we have re-modelled with SEDs. \citet{andre00} proposed that the $M-L$ diagram was a useful diagnostic tool for class 0 and class I low-mass protostars, which provides insight into a source's evolutionary status. \citet{hill09} showed that care should be taken with interpretation of  $M-L$ diagrams drawn from SEDs.

\begin{figure}
 \includegraphics[width=\hsize]{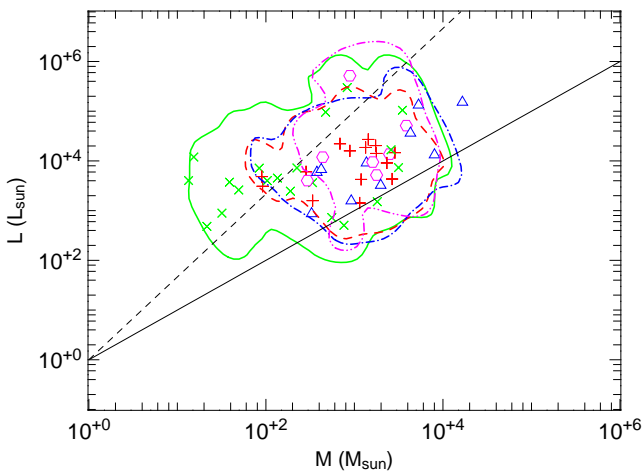}
\caption{Mass-Luminosity diagram for the different classes of source in the sample. Each of the points represents the median value  (\emph{i.e.} the values at which the cumulative probability distribution is 0.5) of the mass and luminosity of the individual sources (green cross: mm, red plus sign: m, blue triangles:mr and pink circles: r).  The contour levels represent the regions enclosing 68 per cent of the Bayesian probability for a given class of sources (green full line: mm, red dashed line: m, blue dot-dash line: mr and pink dot-dot-dash line: r). The full and dashed straight lines depict the $M = L$ and $M = L^{0.6}$ relations, respectively. The same figure can be found in \citet[][Fig. 3]{hill09} for sources which were previously SED modelled. \label{fig:lm}}
\end{figure}

Due to the low number of sources which were re-modelled in each class of source, little distinction can be discerned for the different classes of source nor can proposed evolutionary tracks be drawn from this diagram. The  $M-L$ diagram can be found in Figure~\ref{fig:lm}.

\subsection{Determining a mass for \nh\, sources not SED modelled}

Of the 138 sources deemed to have good \nhone\, fits reliable kinetic temperatures were derived for 128 (see section~\ref{sec:detection}). Of these 128 only 52 \nh\, sources were originally SED modelled by us. Consequently, only 52 \nh\, sources have enough information to draw spectral energy distributions and determine their mass.

Assuming that the \nh\, temperature is equivalent to that of the dust temperature \citep{li03, schnee07}, we can estimate the mass of the remaining sources which have reliable \nh\, and millimetre continuum data. Here we ignore sources which are lacking millimetre continuum or distance information.

The 65 \nh\, sources for which we derive a mass from their \nh\, temperature, according to the procedure outlined in section 4.1, and equation 1, of \citet{hill05},  are presented in Table~\ref{tab:nh3_masses}. Sources which have reliable temperature information and an estimate of the error associated with the temperature have both a minimum and maximum estimation of the mass in columns 3 and 4. Those sources for which a mass was derived from an upper limit temperature value do not have a temperature range and are lacking values in these columns. These masses should be treated with caution and are an estimate of the mass only.

\begin{table}
\caption{Mass of the 65 \nh\, cores that could not be SED modelled. A minimum and maximum mass are determined from the errors on the temperature  (see Table~\ref{tab:nh3_11_22_fits}). For those sources in Rel. group 3 an error on the temperature could not be obtained, and in this instance there is no corresponding error on the mass given in this table. In a few instances, no physically meaningful maximum temperature could be derived, and these sources will not have a minimum mass in this table.}\label{tab:nh3_masses}
\begin{tabular}{lccc}
\hline
Name & M & M$_{\mathrm{min}}$ & M$_{\mathrm{max}}$ \\
\hline  
G269.15-1.13  & 4.0e+02 & 1.8e+01 & 1.0e+03 \\
G291.256-0.769  & 4.6e+02 & 9.0e+01 & 1.6e+03 \\
G291.256-0.743  & 5.3e+02 & 3.4e+02 & 7.2e+02 \\
G291.309-0.681  & 6.1e+02 & 8.1e+01 & 1.4e+03 \\
G291.576-0.468  & 1.1e+03 & -- & -- \\
G291.58-0.53  & 4.3e+03 & -- & -- \\
G294.97-1.7  & 1.5e+02 & -- & -- \\
G305.145+0.208  & 1.3e+02 & -- & -- \\
G305.137+0.069  & 5.3e+02 & -- & -- \\
G305.192-0.006  & 2.3e+02 & 4.5e+01 & 6.4e+02 \\
G305.197+0.007  & 4.6e+02 & 2.3e+02 & 8.6e+02 \\
G305.21+0.21  & 5.3e+02 & 1.2e+01 & 1.3e+03 \\
G305.226+0.275  & 1.1e+02 & 6.7e+01 & 1.8e+02 \\
G305.228+0.286  & 9.3e+00 & 2.5e-01 & 2.3e+01 \\
G305.238+0.261  & 3.1e+02 & 1.2e+02 & 6.6e+02 \\
G305.248+0.245  & 4.6e+02 & -- & -- \\
G305.233-0.023  & 1.0e+02 & 5.5e+01 & 1.8e+02 \\
G305.269-0.010  & 6.1e+02 & 5.1e+02 & 7.6e+02 \\
G305.362+0.185  & 3.5e+02 & 1.4e+02 & 6.2e+02 \\
G305.361+0.151  & 3.1e+02 & -- & -- \\
G305.37+0.21  & 9.3e+02 & -- & -- \\
G305.55+0.01  & 1.1e+02& -- & 3.4e+02 \\
G305.552+0.012  & 2.0e+02 & -- & -- \\
G305.561+0.012  & 2.7e+02 & -- & 6.7e+02 \\
G305.776-0.251  & 1.1e+02 & -- & 5.6e+04 \\
G305.81-0.25  & 9.3e+02 & 3.7e+02 & 1.8e+03 \\
G306.33-0.3  & 1.1e+02 & -- & -- \\
G306.343-0.302  & 3.8e+01 & -- & -- \\
G309.917+0.494  & 1.3e+02 & -- & -- \\
G309.92+0.4  & 1.4e+03 & -- & -- \\
G318.92-0.68  & 1.7e+02 & 4.9e+01 & 3.5e+02 \\
G332.695-0.609  & 1.9e+03 & -- & -- \\
G332.725-0.62  & 3.1e+02 & -- & -- \\
G332.627-0.511  & 1.0e+02 & 3.8e+01 & 1.7e+02 \\
G0.331-0.164  & 2.5e+03 & -- & -- \\
G0.310-0.170  & 5.3e+02 & -- & -- \\
G0.627-0.848  & 5.0e+01 & -- & -- \\
G5.90-0.44  & 3.5e+02 & 1.2e+02 & 7.1e+02 \\
G5.948-1.125  & 6.6e+01 & -- & -- \\
G5.971-1.158  & 8.7e+01 & -- & -- \\
G8.644-0.395  & 2.7e+02 & 1.3e+02 & 4.5e+02 \\
G8.713-0.364  & 1.1e+03 & 7.8e+02 & 1.4e+03 \\
G8.735-0.362  & 1.1e+03 & 5.7e+02 & 1.6e+03 \\
G8.709-0.412  & 4.0e+02 & 2.8e+02 & 7.1e+02 \\
G8.724-0.401  & 2.3e+02 & 1.4e+02 & 4.6e+02 \\
G8.718-0.410  & 8.7e+01 & 6.1e+01 & 1.3e+02 \\
G10.32-0.15  & 1.5e+02 & 3.9e+01 & 3.5e+02 \\
G10.620-0.441  & 3.5e+02 & 2.1e+02 & 6.4e+02 \\
G11.075-0.384  & 1.1e+03 & -- & -- \\
G11.93-0.14  & 3.5e+02 & -- & -- \\
G11.942-0.256  & 2.0e+02 & 7.0e+01 & 4.3e+02 \\
G12.112-0.125  & 3.3e+03 & -- & -- \\
G12.722-0.218  & 6.1e+02 & -- & -- \\
G12.885-0.222  & 1.7e+02 & 9.5e+01 & 2.9e+02 \\
G12.892-0.226  & 1.5e+02 & 1.2e+02 & 1.8e+02 \\
G12.878-0.226  & 2.0e+02 & -- & -- \\
G12.897-0.281  & 8.7e+01 & 4.6e+01 & 1.6e+02 \\
\end{tabular}
\end{table}

\begin{table}
\contcaption{}
\begin{tabular}{lccc}
\hline
Name & M & M$_{\mathrm{min}}$ & M$_{\mathrm{max}}$ \\
\hline  
G12.914-0.280  & 1.7e+02 & -- & -- \\
G12.938-0.272  & 1.5e+02 & -- & -- \\
G11.49-1.48  & 5.3e+02 & -- & -- \\
G16.580-0.079  & 4.0e+02 & -- & -- \\
G22.36+0.07B  & 2.5e+03 & -- & -- \\
G23.43-0.18  & 1.3e+04 & -- & -- \\
G23.409-0.228  & 1.2e+03 & 6.6e+02 & 1.9e+03 \\
G23.754+0.095  & 3.1e+02 & -- & -- \\
\hline
\end{tabular}
\end{table}

Though the mass drawn from this method is derived solely from the millimetre continuum flux, rather than a SED sampled at various wavelengths, it provides a useful approximation to the source mass in the absence of further data. 
\citet{hill05} showed that the contribution of free-free emission to the millimetre flux is expected to be negligible compared with the dust emission. We recognise the potential for contributions from free-free emission that may be missed from the interferometric survey \citep{walsh98} used to draw these conclusions and we caution interpretation of these results  \citep[cf.][]{longmore09}.

%_______________________________________________________

\section{Discussion}

\subsection{Physical properties of the clumps}

Previous work has shown the MM-only cores of our sample to be excellent candidates for young massive protostars. Their characteristic physical properties, such as temperature and mass, suggested that these objects are indicative of the earliest stages of massive star formation. These conclusions, which were drawn from (sub)millimetre continuum emission observations and spectral energy distribution modelling, are not without their caveats. SED modelling is subject to ambiguities and is heavily reliant on well constrained temperatures in order to produce a reliable mass estimate. In addition, it is not yet known whether the MM-only core is indeed forming massive stars.

Using the kinetic temperatures as derived from this \nh\, line study, we  are now in a position to ascertain which physical properties of the MM-only cores are similar to those of sources with known star formation activity, and which physical properties differ.

% Differences
Analysis, in the previous sections, has revealed the MM-only class to have smaller \nh\, linewidths on average than class M, MR and R - those sources with a methanol maser and/or radio continuum association. This is true for both ammonia transitions. SED modelling, using the Bayesian inference method, in conjunction with well constrained kinetic temperatures, as derived from ammonia, reveals the MM-only cores to be the least massive of the sample. 

% Common points
Interestingly, class MM sources display many similarities in the distributions of the other physical parameters compared with the other classes of source. They have similar luminosities, \nh\, flux densities hence brightness level, and column densities as the other classes. The simultaneous analysis of the \nh\, and continuum data shows that the distribution of the temperature (Fig.~\ref{fig:sedcumul}) for the MM-only sources is not very distinct from that of the other classes, i.e., those with known star formation activity, but seems to extends to slightly smaller temperatures.

\subsection{Are they gravitationally bound?}

From SED modelling, it is clear that both \nh\, and continuum observations provide complementary constraints on SED derived source parameters. Even if each of these constraints, taken individually, provide limited information on the nature of the cores, the combination of these constraints reveal several trends that give insight into the cores and their evolutionary status: Class MM are smaller \citep{hill05}, less massive and more quiescent (smaller turbulent linewidths) than cores with signatures of star formation.

\begin{figure*}
  \includegraphics[angle=270,width=\hsize]{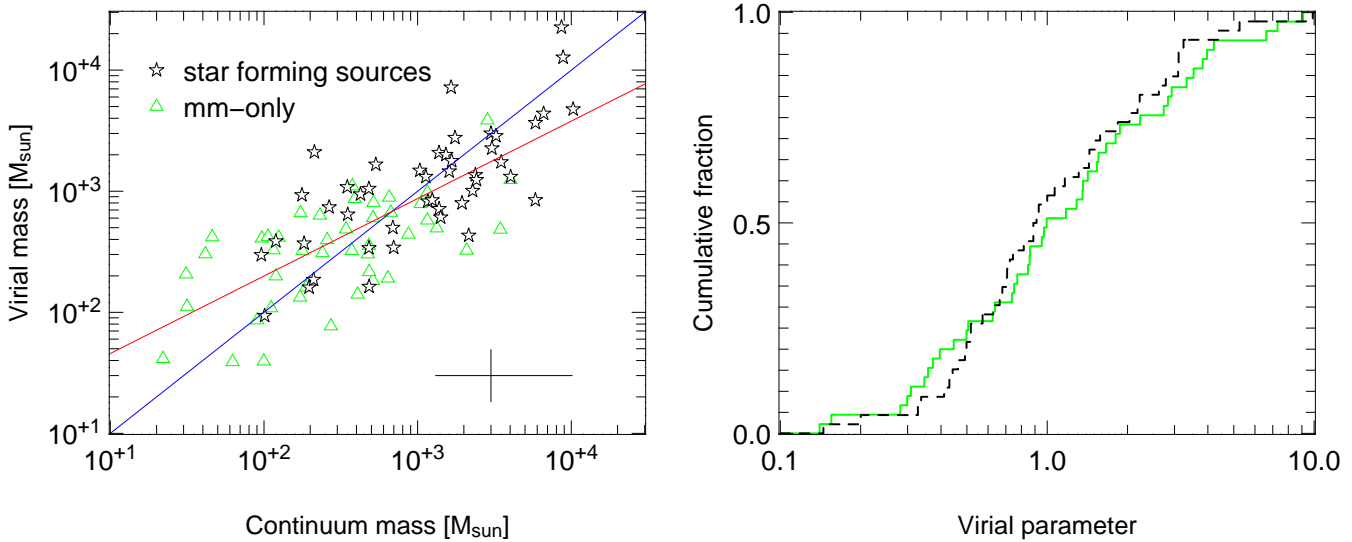}
  \caption{Left: The virial mass of the source as a function of the continuum mass. The blue line is where the two masses are equal, whilst the red line is the fit to the data, which has a correlation co-efficient of 0.74. The average error bars are as indicated on the plot, assuming a conservative 50 per cent error on the radius. Right: The cumulative distribution of the virial parameter. Same colour scheme as used in the left-hand plot.
  \label{fig:virial_mass}}
\end{figure*}

With this information,  we can ascertain the gravitational equilibrium of the various classes of source in the sample. A plot presenting the virial mass $  M_\mathrm{virial} = \Delta V^2\,R / G  \label{eq:virial_mass} $, as a function of the gas mass derived from continuum observations (i.e. SIMBA), $M_\mathrm{continuum}$, is presented in Fig.~\ref{fig:virial_mass}. This plot contains all sources which have a reliability flag (Rel. Group) of 1 or 2 in Table~\ref{tab:nh3_derivs}, though as a check, by including the sources with a reliability flag of 3, the correlation coefficient and the line of best fit do not differ significantly.

As can be seen in Fig~\ref{fig:virial_mass}, there is a strong correlation between the virial mass and gas mass of a source, with a correlation coefficient of 0.74. As the continuum mass of a source increases so too does the virial mass. Interestingly, the distribution of the virial parameter is very similar for the MM-only sources and the active star formation sample (class M+MR+R) suggesting a very similar gravitational state, despite the lower masses of the MM-only sources.

The virial parameter  $\alpha =M_\mathrm{virial}/M_\mathrm{continuum}$ \citep{bertoldi92} has a median value of 0.95 (see Fig.~\ref{fig:virial_mass}, right panel), suggesting that the clumps are virialised. The precise value must be interpreted with caution given the uncertainties on both the virial mass (in particular on the size of the clumps) and on the continuum mass (e.g. uncertain dust emissivities), but it is clear that the clumps, on average, do not differ significantly from being virialised. 

There is however a distribution of the virial parameter as a function of the source mass, ranging from $\approx 0.1$ to $\approx 10$. The virial mass appears to vary as $M_\mathrm{virial} \propto M_\mathrm{continuum}^{0.6}$, indicating that the more massive clumps are slightly more gravitationally bounded. This translates as a slightly larger median virial parameter of 1 for the MM-only cores, compared with the star forming sources (median $\alpha=0.9$).

\begin{figure}
 \includegraphics[width=\hsize]{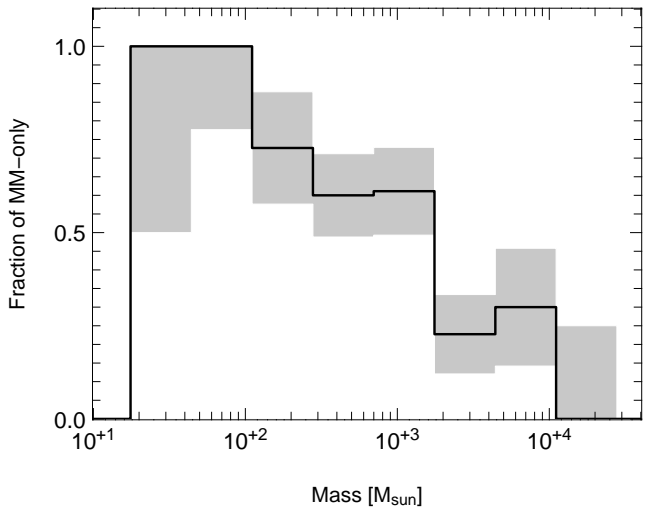}
 \caption{Fraction of MM-only sources as a function of the mass of the object. The error bars are identified by the grey boxes. \label{fig:mm:cont}}
\end{figure}

\subsection{MM-only sources in the context of massive star formation}

High-mass prestellar cores, or high-mass starless cores, the precursors of high-mass protostars in an analogous scaled-up version of low-mass starless cores, are still a missing piece of the high-mass star formation puzzle. Whether they may exist will serve to constrain competing massive star formation theories which propose coalescence and  accretion as possible formation mechanisms. Infrared dark clouds \citep{jackson08} have been proposed as possible candidates to host such starless objects. However, high-angular resolution observations of a small sample of candidate massive starless cores have systematically revealed embedded, accreting protostars inside \citep[e.g.][]{pillai06}.

We have identified a new class of source, the MM-only core, which are excellent candidates for early stage protostars or massive young stellar objects. SED analysis here reveals the MM-only sources to have a similar mass range, with a larger number of less massive sources, compared to the star forming sources, and very similar physical properties as these sources. The similar distribution of the virial state of equilibrium suggests that the MM-only sources are just as likely to form stars. These results 
 suggest that the MM-only sources may in fact be scaled down objects compared with the star forming sources. The smaller mass, radii and linewidths of the MM-only cores further suggests that they will possibly form intermediate mass stars rather than high-mass stars.

Alternatively, the MM-only cores may not produce stars, and could remain starless transient clumps \citep[e.g.][]{beuther09, buckle09}. Further spectral line observations designed to ascertain the chemistry of the MM-only core are currently underway.

\subsection{Is the MM-only population comprised of sub-samples?}\label{sec:subsamples}

It is also possible that the MM-only clumps could be comprised of multiple sub-samples or populations, with one interpretation that each of these sub-populations likely follow different evolutionary paths. In this instance, they could be comprised of both cores with high and low mass stars, cores which are in the very earliest stages of evolution, and also possibly some cores that  are not destined to begat high mass stars.

Attempts to break class MM sources into multiple populations with respect to the mass, radius, linewidth, luminosity and temperature reveal the MM-only population to have a continuous distribution with respect to each of these parameters (see for instance Fig.~\ref{fig:sedcumul}, \ref{fig:lm}  and \ref{fig:virial_mass}).  Consequently, there is currently no physical indication that the MM-only sources are comprised of distinct sub-populations.

Despite this, from our results we can infer that the MM-only sources will have different star formation histories depending on their initial mass. Figure~\ref{fig:mm:cont} presents a plot of the fraction of MM-only sources as a function of the mass of the clump. All of the sources in our sample less than 100\mstar, are MM-only sources, and the fraction of objects that do not show any signature of star formation decreases as the mass of the clump increases.

This indicates that the most massive clumps have a higher efficiency to form massive stars. Additionally it may suggest that there is a dichotomy in mass around 100--200\mstar\, for the MM-only sample. The MM-only sources below this threshold  will likely not form massive stars.

Assuming an initial mass function (IMF) as per \citet{chabrier03}, and a star formation efficiency of 50 per cent, then we would need a minimum clump mass of $\approx$\,500\mstar\, in order to support the formation of one star in excess of 10\mstar. That is, the maximum mass of the original clump dictates the size of stars that can form. Thus, clumps with masses lower than 100\mstar\, will not form massive stars, whilst more massive MM-only sources on the contrary have enough material to form stars in excess of 10\mstar. 

Assuming that the evolution of protostellar objects is faster for more massive objects could explain why we detect a smaller fraction of MM-only sources i.e. before signatures of star formation, for increasingly massive objects.  In this instance, the MM-only cores do represent targets of choice to assess the earliest stages of star formation.

%________________________________________________________
\section{Conclusions}

We have undertaken an ammonia molecular line study, in the lowest two inversion transitions, of young massive star formation regions in the southern hemisphere as part of an effort to characterise the earliest stages of high-mass star formation. The sample targeted was derived from the millimetre continuum emission studies of \citet{hill05, hill06} and included sources with and without signatures of high mass star formation. In total, 244 sources were observed in both ammonia transitions, using the Parkes radio telescope. Of these, 138 had detections ($>$3-$\sigma$) in \nhone, including two sources with two velocity components, and 102 in \nhtwo. The MM-only sources are not more or less likely to be detected in either transition than the sources with a methanol maser and/or radio continuum association.

The spectral line properties of the sources have been determined from gaussian fits to the lines: linewidth, flux density, opacity. In addition, physical parameters, such as the rotational and kinetic temperatures as well as the column density, were derived from the spectra. From the kinetic temperature, and revisiting previous SED modelling techniques \citep{hill09}, we have determined the mass and luminosity of the sources in the sample. We have used the Bayesian inference method of SED modelling, which provides robust estimates of the parameters as well as good estimates of the uncertainties associated with each parameter, further allowing robust statistical conclusions.

Combining continuum and line observations has proven to be quite powerful. In our case, ammonia observations and SED modelling are very complementary in terms of parameter determinations. It is clear, that the kinetic temperature, as derived from ammonia, used in combination with SED modelling has constrained the range of validity for the temperature, which in turn has constrained the range of validity for the mass and luminosity for each of the sources in our sample.

The MM-only sources, those with no overt signs of massive star formation, have smaller \nhone\, and (2,2) linewidths on average compared with sources associated with a methanol maser and/or radio continuum source. They are also less massive and smaller on average but reach the same upper values for both parameters. Because the MM-only sources have  similar brightness/flux, column densities and similar temperatures as star forming sources, they have the potential to form stars. The least massive MM-only sources, cannot form high mass stars and will likely proceed to form intermediate mass stars, whilst the more massive clumps, are more interestingly  strong candidates for early stage massive protostars. The different hypotheses will result in different internal structures. 

Higher spatial resolution observations with ALMA will allow resolution of the internal structures of the MM-only cores, which may then allow us to distinguish between starless cores that will begat massive stars and those that are transient.

\vspace{-0.2cm}
\section*{Acknowledgments}

The authors would like to thank S.~Mader for observational support and J.~Reynolds \& E. Carretti for help with calibration of the observations and derivation of the gain elevation correction. We also extend thanks to an anonymous referee whose comments served to improve the manuscript. C.~Pinte acknowledges funding support of the European Commission’s Seventh Framework Program as a Marie Curie Intra-European Fellow (PIEF-GA-2008-220891). The Parkes telescope is part of the Australia Telescope which is funded by the Commonwealth of Australia for operation as a National Facility managed by CSIRO.

\vspace{-0.2cm}
\bibliographystyle{mn2e}
%
%\bibliography{/Users/thill/data/bib/aa,/Users/thill/data/bib/references}
%\bibliography{references}
\bibliography{/Volumes/home/thill/data/bib/aa,/Volumes/home/thill/data/bib/references} %laptop

\begin{thebibliography}{}

\bibitem[\protect\citeauthoryear{{Andr\'e}, {Ward-Thompson} \&
  {Barsony}}{{Andr\'e} et~al.}{2000}]{andre00}
{Andr\'e} P.,  {Ward-Thompson} D.,    {Barsony} M.,  2000, Protostars and
  Planets IV, p.~59

\bibitem[\protect\citeauthoryear{{Bertoldi} \& {McKee}}{{Bertoldi} \&
  {McKee}}{1992}]{bertoldi92}
{Bertoldi} F.,  {McKee} C.~F.,  1992, ApJ, 395, 140

\bibitem[\protect\citeauthoryear{{Beuther}, {Churchwell}, {McKee} \&
  {Tan}}{{Beuther} et~al.}{2007}]{beuther07}
{Beuther} H.,  {Churchwell} E.~B.,  {McKee} C.~F.,    {Tan} J.~C.,  2007, in
  {Reipurth} B.,  {Jewitt} D.,   {Keil} K.,  eds, Protostars and Planets V {The
  Formation of Massive Stars}.
pp 165--180

\bibitem[\protect\citeauthoryear{{Beuther} \& {Henning}}{{Beuther} \&
  {Henning}}{2009}]{beuther09}
{Beuther} H.,  {Henning} T.,  2009, A\&A, 503, 859

\bibitem[\protect\citeauthoryear{{Beuther}, {Schilke}, {Menten}, {Motte},
  {Sridharan} \& {Wyrowski}}{{Beuther} et~al.}{2002}]{beuther02}
{Beuther} H.,  {Schilke} P.,  {Menten} K.~M.,  {Motte} F.,  {Sridharan} T.~K.,
    {Wyrowski} F.,  2002, ApJ, 566, 945

\bibitem[\protect\citeauthoryear{{Bonnell}, {Bate} \& {Zinnecker}}{{Bonnell}
  et~al.}{1998}]{bonnell98}
{Bonnell} I.~A.,  {Bate} M.~R.,    {Zinnecker} H.,  1998, MNRAS, 298, 93

\bibitem[\protect\citeauthoryear{{Buckle et al.}}{{Buckle et
  al.}}{2009}]{buckle09}
{Buckle et al.} J.~V.,  2009, ArXiv e-prints

\bibitem[\protect\citeauthoryear{{Chabrier}}{{Chabrier}}{2003}]{chabrier03}
{Chabrier} G.,  2003, ApJL, 586, L133

\bibitem[\protect\citeauthoryear{{Churchwell}, {Walmsley} \&
  {Cesaroni}}{{Churchwell} et~al.}{1990}]{churchwell90b}
{Churchwell} E.,  {Walmsley} C.~M.,    {Cesaroni} R.,  1990, A\&AS, 83, 119

\bibitem[\protect\citeauthoryear{{Danby}, {Flower}, {Valiron}, {Schilke} \&
  {Walmsley}}{{Danby} et~al.}{1988}]{danby88}
{Danby} G.,  {Flower} D.~R.,  {Valiron} P.,  {Schilke} P.,    {Walmsley} C.~M.,
   1988, MNRAS, 235, 229

\bibitem[\protect\citeauthoryear{{Evans}, {Shirley}, {Mueller} \&
  {Knez}}{{Evans} et~al.}{2002}]{evans02}
{Evans} N.~J.,  {Shirley} Y.~L.,  {Mueller} K.~E.,    {Knez} C.,  2002, in
  {Crowther} P.,  ed., Hot Star Workshop III: The Earliest Phases of Massive
  Star Birth Vol.~267 of Astronomical Society of the Pacific Conference Series,
  {The Formation and Early Evolution of Massive Stars}.
pp 17--+

\bibitem[\protect\citeauthoryear{{Garay} \& {Lizano}}{{Garay} \&
  {Lizano}}{1999}]{garay99}
{Garay} G.,  {Lizano} S.,  1999, PASP, 111, 1049

\bibitem[\protect\citeauthoryear{{Hill}, {Burton}, {Minier}, {Thompson},
  {Walsh}, {Hunt-Cunningham} \& {Garay}}{{Hill} et~al.}{2005}]{hill05}
{Hill} T.,  {Burton} M.~G.,  {Minier} V.,  {Thompson} M.~A.,  {Walsh} A.~J.,
  {Hunt-Cunningham} M.,    {Garay} G.,  2005, MNRAS, 363, 405

\bibitem[\protect\citeauthoryear{{Hill}, {Pinte}, {Minier}, {Burton} \&
  {Cunningham}}{{Hill} et~al.}{2009}]{hill09}
{Hill} T.,  {Pinte} C.,  {Minier} V.,  {Burton} M.~G.,    {Cunningham} M.~R.,
  2009, MNRAS, 392, 768

\bibitem[\protect\citeauthoryear{{Hill}, {Thompson}, {Burton}, {Walsh},
  {Minier}, {Cunningham} \& {Pierce-Price}}{{Hill} et~al.}{2006}]{hill06}
{Hill} T.,  {Thompson} M.~A.,  {Burton} M.~G.,  {Walsh} A.~J.,  {Minier} V.,
  {Cunningham} M.~R.,    {Pierce-Price} D.,  2006, MNRAS, 368, 1223

\bibitem[\protect\citeauthoryear{{Ho} \& {Townes}}{{Ho} \&
  {Townes}}{1983}]{ho83}
{Ho} P.~T.~P.,  {Townes} C.~H.,  1983, Ann. Rev. Astr. Ap., 21, 239

\bibitem[\protect\citeauthoryear{{Jackson}, {Finn}, {Rathborne}, {Chambers} \&
  {Simon}}{{Jackson} et~al.}{2008}]{jackson08}
{Jackson} J.~M.,  {Finn} S.~C.,  {Rathborne} J.~M.,  {Chambers} E.~T.,
  {Simon} R.,  2008, ApJ, 680, 349

\bibitem[\protect\citeauthoryear{{Jijina}, {Myers} \& {Adams}}{{Jijina}
  et~al.}{1999}]{jijina99}
{Jijina} J.,  {Myers} P.~C.,    {Adams} F.~C.,  1999, Astrophys. J. Supp.
  Series, 125, 161

\bibitem[\protect\citeauthoryear{{Kruegel} \& {Walmsley}}{{Kruegel} \&
  {Walmsley}}{1984}]{kruegel84}
{Kruegel} E.,  {Walmsley} C.~M.,  1984, A\&A, 130, 5

\bibitem[\protect\citeauthoryear{{Krumholz}, {Klein}, {McKee}, {Offner} \&
  {Cunningham}}{{Krumholz} et~al.}{2009}]{krumholz09}
{Krumholz} M.~R.,  {Klein} R.~I.,  {McKee} C.~F.,  {Offner} S.~S.~R.,
  {Cunningham} A.~J.,  2009, Science, 323, 754

\bibitem[\protect\citeauthoryear{{Li}, {Goldsmith} \& {Menten}}{{Li}
  et~al.}{2003}]{li03}
{Li} D.,  {Goldsmith} P.~F.,    {Menten} K.,  2003, ApJ, 587, 262

\bibitem[\protect\citeauthoryear{{Longmore}, {Burton}, {Barnes}, {Wong},
  {Purcell} \& {Ott}}{{Longmore} et~al.}{2007}]{longmore07}
{Longmore} S.~N.,  {Burton} M.~G.,  {Barnes} P.~J.,  {Wong} T.,  {Purcell}
  C.~R.,    {Ott} J.,  2007, MNRAS, 379, 535

\bibitem[\protect\citeauthoryear{{Longmore}, {Burton}, {Keto}, {Kurtz} \&
  {Walsh}}{{Longmore} et~al.}{2009}]{longmore09}
{Longmore} S.~N.,  {Burton} M.~G.,  {Keto} E.,  {Kurtz} S.,    {Walsh} A.~J.,
  2009, MNRAS, pp 1151--+

\bibitem[\protect\citeauthoryear{{Longmore}, {Burton}, {Minier} \&
  {Walsh}}{{Longmore} et~al.}{2006}]{longmore06}
{Longmore} S.~N.,  {Burton} M.~G.,  {Minier} V.,    {Walsh} A.~J.,  2006,
  MNRAS, 369, 1196

\bibitem[\protect\citeauthoryear{{Longmore}, {Burton}, {Purcell}, {Barnes} \&
  {Ott}}{{Longmore} et~al.}{2008}]{longmore08}
{Longmore} S.~N.,  {Burton} M.~G.,  {Purcell} C.~R.,  {Barnes} P.,    {Ott} J.,
   2008, in {Beuther} H.,  {Linz} H.,   {Henning} T.,  eds, Massive Star
  Formation: Observations Confront Theory Vol.~387 of Astronomical Society of
  the Pacific Conference Series, {Determining the Relative Evolutionary Stages
  of Very Young Massive Star Formation Regions}.
pp 58--64

\bibitem[\protect\citeauthoryear{{McKee} \& {Tan}}{{McKee} \&
  {Tan}}{2003}]{mckee03}
{McKee} C.~F.,  {Tan} J.~C.,  2003, ApJ, 585, 850

\bibitem[\protect\citeauthoryear{{Menten}, {Pillai} \& {Wyrowski}}{{Menten}
  et~al.}{2005}]{menten05}
{Menten} K.~M.,  {Pillai} T.,    {Wyrowski} F.,  2005, in IAU Symposium
  {Initial conditions for massive star birth-Infrared dark clouds}.
pp 23--34

\bibitem[\protect\citeauthoryear{{Minier}, {Booth} \& {Conway}}{{Minier}
  et~al.}{2000}]{minier00}
{Minier} V.,  {Booth} R.~S.,    {Conway} J.~E.,  2000, A\&A, 362, 1093

\bibitem[\protect\citeauthoryear{{Minier}, {Burton}, {Hill}, {Pestalozzi},
  {Purcell}, {Garay}, {Walsh} \& {Longmore}}{{Minier} et~al.}{2005}]{minier05}
{Minier} V.,  {Burton} M.~G.,  {Hill} T.,  {Pestalozzi} M.~R.,  {Purcell}
  C.~R.,  {Garay} G.,  {Walsh} A.~J.,    {Longmore} S.,  2005, A\&A, 429, 945

\bibitem[\protect\citeauthoryear{{Molinari}, {Brand}, {Cesaroni} \&
  {Palla}}{{Molinari} et~al.}{1996}]{molinari96}
{Molinari} S.,  {Brand} J.,  {Cesaroni} R.,    {Palla} F.,  1996, A\&A, 308,
  573

\bibitem[\protect\citeauthoryear{{Pestalozzi}, {Minier} \&
  {Booth}}{{Pestalozzi} et~al.}{2005}]{pestalozzi05}
{Pestalozzi} M.~R.,  {Minier} V.,    {Booth} R.~S.,  2005, A\&A, 432, 737

\bibitem[\protect\citeauthoryear{{Pillai}, {Wyrowski}, {Carey} \&
  {Menten}}{{Pillai} et~al.}{2006}]{pillai06}
{Pillai} T.,  {Wyrowski} F.,  {Carey} S.~J.,    {Menten} K.~M.,  2006, A\&A,
  450, 569

\bibitem[\protect\citeauthoryear{{Pinte et al.}}{{Pinte et
  al.}}{2008}]{pinte08}
{Pinte et al.} C.,  2008, A\&A, 489, 633

\bibitem[\protect\citeauthoryear{{Robitaille}, {Whitney}, {Indebetouw}, {Wood}
  \& {Denzmore}}{{Robitaille} et~al.}{2006}]{robitaille06}
{Robitaille} T.~P.,  {Whitney} B.~A.,  {Indebetouw} R.,  {Wood} K.,
  {Denzmore} P.,  2006, Astrophys. J. Supp. Series, 167, 256

\bibitem[\protect\citeauthoryear{{Schnee} \& {Sargent}}{{Schnee} \&
  {Sargent}}{2007}]{schnee07}
{Schnee} S.,  {Sargent} A.,  2007, in Bulletin of the American Astronomical
  Society Vol.~38 of Bulletin of the American Astronomical Society, {Dust and
  Gas Temperature of the Prestellar Core TMC-1C}.
pp 879--

\bibitem[\protect\citeauthoryear{{Sridharan}, {Beuther}, {Saito}, {Wyrowski} \&
  {Schilke}}{{Sridharan} et~al.}{2005}]{sridharan05}
{Sridharan} T.~K.,  {Beuther} H.,  {Saito} M.,  {Wyrowski} F.,    {Schilke} P.,
   2005, ApJL, 634, L57

\bibitem[\protect\citeauthoryear{{Sridharan}, {Beuther}, {Schilke}, {Menten} \&
  {Wyrowski}}{{Sridharan} et~al.}{2002}]{sridharan02}
{Sridharan} T.~K.,  {Beuther} H.,  {Schilke} P.,  {Menten} K.~M.,    {Wyrowski}
  F.,  2002, ApJ, 566, 931

\bibitem[\protect\citeauthoryear{{Tafalla}, {Myers}, {Caselli} \&
  {Walmsley}}{{Tafalla} et~al.}{2004}]{tafalla04}
{Tafalla} M.,  {Myers} P.~C.,  {Caselli} P.,    {Walmsley} C.~M.,  2004, A\&A,
  416, 191

\bibitem[\protect\citeauthoryear{{Ungerechts}, {Winnewisser} \&
  {Walmsley}}{{Ungerechts} et~al.}{1986}]{ungerechts86}
{Ungerechts} H.,  {Winnewisser} G.,    {Walmsley} C.~M.,  1986, A\&A, 157, 207

\bibitem[\protect\citeauthoryear{{Walsh}, {Burton}, {Hyland} \&
  {Robinson}}{{Walsh} et~al.}{1998}]{walsh98}
{Walsh} A.~J.,  {Burton} M.~G.,  {Hyland} A.~R.,    {Robinson} G.,  1998,
  MNRAS, 301, 640

\bibitem[\protect\citeauthoryear{{Whitney}, {Wood}, {Bjorkman} \&
  {Cohen}}{{Whitney} et~al.}{2003}]{whitney03}
{Whitney} B.~A.,  {Wood} K.,  {Bjorkman} J.~E.,    {Cohen} M.,  2003, ApJ, 598,
  1079

\bibitem[\protect\citeauthoryear{{Williams}, {Fuller} \&
  {Sridharan}}{{Williams} et~al.}{2004}]{williams04}
{Williams} S.~J.,  {Fuller} G.~A.,    {Sridharan} T.~K.,  2004, A\&A, 417, 115

\bibitem[\protect\citeauthoryear{{Zinnecker} \& {Yorke}}{{Zinnecker} \&
  {Yorke}}{2007}]{zinnecker07}
{Zinnecker} H.,  {Yorke} H.~W.,  2007, Ann. Rev. Astr. Ap., 45, 481

\end{thebibliography}
\expandafter\ifx\csname natexlab\endcsname\relax\def\natexlab#1{#1}\fi
%need for mn2e to work properly

\vspace{-0.2cm}
\appendix
\section{\nhone\, and (2,2) Spectra}

The \nhone\, and (2,2) inversion spectra detected by the K-band receiver on the Parkes Telescope are presented here for all sources, including those in the sample figure (Fig.~\ref{fig:sample:spectra}). Sources are presented in right ascension order, consistent with Table ~\ref{tab:nh3_11_22_fits} and the complementary continuum images of \citet{hill05}. The x-axis is in units of velocity  (\kms) whilst the y-axis is the flux density (Jy/beam) or strength of the detection. Both the \vlsr\, and flux density of the sources are presented in Table~\ref{tab:nh3_11_22_fits}. The fits as returned from CLASS are overlaid on the spectra in green. 

%\input spectra_print.tex

%\section{ONLINE MATERIAL}

%\input spectra_online.tex

\bsp

\label{lastpage}

\end{document}